\documentclass[fleqn]{article}

\usepackage{graphicx}% Include figure files
\usepackage{dcolumn}% Align table columns on decimal point
\usepackage{bm}% bold math
\usepackage{amsmath}
\usepackage{amsfonts}
\usepackage{subfigure}

\def\abstract#1{\vskip 7mm 
        \begin{center}{\large Abstract}\par \smallskip
                \begin{minipage}[c]{12cm}
                        \small #1
                \end{minipage}
        \end{center}
}
\def\title#1{\begin{center}{\Large\bf #1}\end{center}}
\def\author#1{\vskip 5mm \begin{center}{#1}\end{center}}
\def\address#1{\begin{center}{\it #1}\end{center}}
\makeatletter
\@ifundefined{lesssim}{}{}
\@ifundefined{gtrsim}{}{}
\def\vereq#1#2{\lower3pt\vbox{\baselineskip1.5pt \lineskip1.5pt
\ialign{$\m@th#1\hfill##\hfil$\crcr#2\crcr\sim\crcr}}}
\makeatother

\begin{document}

\begin{flushright}
IFT-UAM/CSIC-08-05
\end{flushright}

\vspace{5mm}

\title{%
  A new gravitational wave background from the Big Bang
  }
\author{%
  Juan Garc\'\i a-Bellido\footnote{E-mail: juan.garciabellido@uam.es}\,\, 
  and\,\, Daniel G. Figueroa\footnote{E-mail: daniel.figueroa@uam.es}
  }
\address{%
  Instituto de F\'\i sica Te\'orica CSIC-UAM, \\Universidad Aut\'onoma de Madrid, \\
 Cantoblanco 28049 Madrid, Spain
}

\vspace*{-2mm}

\abstract{\vspace*{2mm}
  The reheating of the universe after hybrid inflation proceeds through
the nucleation and subsequent collision of large concentrations of
energy density in the form of bubble-like structures moving at
relativistic speeds. This generates a significant fraction of energy
in the form of a stochastic background of gravitational waves, whose
time evolution is determined by the successive stages of
reheating: First, tachyonic preheating makes the amplitude of gravity
waves grow exponentially fast. Second, bubble collisions add a new
burst of gravitational radiation. 
Third, turbulent motions finally sets the end of gravitational waves 
production. From then on, these waves propagate unimpeded to us.
We find that the fraction of energy density today in these
primordial gravitational waves could be significant for GUT-scale
models of inflation, although well beyond the frequency range
sensitivity of gravitational wave observatories like LIGO, LISA or
BBO. However, low-scale models could still produce a detectable signal
at frequencies accessible to BBO or DECIGO. For comparison, we have
also computed the analogous %gravitational wave 
background from some chaotic inflation models and 
obtained similar results to those %found 
of other groups. The discovery of such a background would open 
a new observational window into the very early universe, where the details 
of the process of reheating, i.e. the Big Bang, could be explored. 
Thus, it could also serve %in the future 
as a new experimental tool for testing the Inflationary Paradigm.
}

\section{Introduction}

Gravitational waves (GW) are ripples in space-time that travel at the
speed of light, and whose emission by relativistic bodies represents a
robust prediction of General Relativity. % The change in the orbital
%period of a binary pulsar known as PSR 1913+16 was used by Hulse and
%Taylor~\cite{HulseTaylor} to obtain indirect evidence of their
%existence. 
%Although gravitational radiation has not been directly
%detected yet, 
Theoretically, it is expected that the present universe should be
permeated by a diffuse background of GW of either an astrophysical or
cosmological origin~\cite{Maggiore}. %a variety of origins
% either an astrophysical or cosmological origin
%~\cite{Maggiore}. %Astrophysical sources, like the
%gravitational collapse of supernovae or the neutron star and black
%hole binaries' coalescence, produce a stochastic gravitational wave
%background (GWB) which can be understood as coming from unresolved
%point sources. On the other hand, 
%Among the hypothetical backgrounds expected from the early universe, 
%we can cite, for example, the approximately scale-invariant
%background produced during inflation~\cite{Starobinsky}, or the GWB
%generated at  early universe thermal phase transitions.% or
%from relativistic motions of turbulent plasmas.% or from the decay of
%cosmic strings~\cite{Maggiore}.
Fortunately, these backgrounds have
very different spectral signatures 
%shapes and amplitudes 
that might, in the future, 
allow gravitational wave observatories like %the Laser
%Interferometer Gravitational Wave Observatory (
LIGO~\cite{LIGO}, 
%the Laser Interferometer Space Antenna (%
LISA~\cite{LISA}, %the Big Bang Observer (
BBO~\cite{BBO} or %the Decihertz Interferometer
%Gravitational Wave Observatory (
DECIGO~\cite{DECIGO}, to disentangle
their origin~\cite{Maggiore}. Unfortunately, the weakness of
gravity will make this task extremely difficult, requiring a very high
accuracy in order to distinguish one background from another. %It is 
%thus important to characterize as many different sources of GW as
%possible.

There are, indeed, a series of constraints on some of these
backgrounds, coming %the most stringent one coming 
from the %large-scale polarization 
anisotropies in the Cosmic Microwave Background (CMB)~\cite{Elena},
%which might be measured by Planck~\cite{Planck} if the scale of
%inflation is sufficiently high. There are also constraints coming from
from Big Bang nucleosynthesis~\cite{BBN} %since such a background would
%contribute as a relativistic species to the expansion of the universe
%and thus increase the light element abundance. There is also a
%constraint coming 
%\,, 
or from millisecond pulsar timing~\cite{pulsar}.  
%or furthermore, it has recently been proposed a new constraint on a GWB
%even from from CMB anisotropies~\cite{Elena}. 
Most of these constraints
come at very low frequencies, %typically 
from $10^{-18}$ Hz to
$10^{-8}$ Hz, while present and future GW detectors (will) work at 
frequencies of order %1-100 Hz, and planned observatories will range 
%from $10^{-3}$ Hz of LISA to $10^{3}$ Hz of Advanced-LIGO~\cite{Maggiore,LIGO}.
$10^{-3}-10^3$ Hz. If early universe first order phase
transitions~\cite{KosowskyTurner,Nicolis} or cosmic
turbulence~\cite{Turbulence} occurred around the electro-weak (EW)
scale, %there is a chance than the 
GW detectors could have a chance to measure the
corresponding associated backgrounds. However, if those processes 
occurred at the GUT scale, their corresponding backgrounds
will go undetected by the actual detectors, since these cannot reach
the required sensitivity in the high frequency range of $10^7-10^9$
Hz. There are however recent proposals to cover this
range~\cite{Nishizawa2007,Picasso}, which may become competitive in 
the not so far future.

%Present 
Cosmological observations %of the CMB anisotropies and
%the Large Scale Structure (LSS) distribution of matter %
seem to suggest
that something like Inflation must have occurred in the very early
universe. %We ignore what drove inflation and at what scale it took
%place. However, 
Approximately scale-invariant density perturbations,
sourced by quantum fluctuations during inflation, seem to be the most
satisfying explanation for the CMB anisotropies. Together with such
scalar perturbations one also expects tensor perturbations (GW) to be
produced, with an almost scale-free power
spectrum~\cite{Starobinsky}. %Because of the weakness of gravity, this
%primordial inflationary GWB should decouple from the rest of matter as
%soon as it is produced, and move freely through the Universe till
%today. 
%At present, the biggest efforts employed in the search for
%these primordial GW come from the indirect effect that this
%background has on the B-mode polarization anisotropies of the
%CMB~\cite{Planck}, rather than via direct detection. 
The detection of such a background is crucial for early universe 
cosmology because it would help to determine the absolute energy 
scale of inflation, a quantity that for the moment is still 
uncertain, and would open the exploration of physics at very high energies.

However, in the early universe, after inflation, other GWB
could have been produced at shorter wavelengths, in a more 'classical'
manner rather than sourced by quantum fluctuations. In particular,
whenever there are large and fast moving inhomogeneities in a matter
distribution, one expects the emission of GW. % This is much like the
%situation in classical electrodynamics, but with some differences. 
At large distances from a source, %the amplitude of the electromagnetic
%fieldis expressed as $A_i\simeq \dot d_i /cr$, %the first derivative of the 
%with $d_i$ the dipole moment of the charge distribution %of the source,
%, while 
the amplitude of the GW is given by %the second derivative
%of the quadrupole moment of the mass distribution, 
$h_{ij} \simeq
G\ddot Q_{ij} / c^4r$, with $Q_{ij}$ the quadrupole moment of the 
mass distribution. %In both cases, 
The larger the velocity of the
matter distribution, the larger the amplitude of the radiation
produced. %Nevertheless, the main difference between the two cases is
However, because of the weakness of gravity, %to that of
%electromagnetism, makes a crucial difference: 
in order to produce a significant amount of
gravitational radiation, it is required %that the motion of huge masses
%occurs at speeds %close to that of light for the case of astrophysical
%sources, or 
a very relativistic motion (and high density contrasts) in the matter 
distribution of a source.
%for the case of cosmological sources.
Fortunatelly, this is indeed believed to be the situation at the end of inflaton,
during the conversion of the huge energy density driving inflation
into radiation and matter, at the so-called {\it reheating} of the
Universe~\cite{preheating}, i.e. at the Big Bang.
% Such an event corresponds to 
%the actual Big Bang of the Standard Cosmological Model.

Note that any background of GW coming from the early universe, if
generated below Planck scale, immediately decoupled upon production
%as can be easily understood: %by the following dimensional analysis
%argument. 
%Assuming that gravitons were in thermal equilibrium with the
%early universe plasma at a temperature $T$, the gravitons' cross
%section should be of order $\sigma\sim G^2T^2$. Then, given the
%graviton number density $n\sim T^3$ and velocity $v = 1$, the
%gravitons' interaction rate should be $\Gamma = \left\langle n\sigma
%v\right\rangle\,\sim\,T^5/M_p^4$. Since the Hubble rate is $H\sim
%T^2/M_p$, then $\Gamma\,\sim\,H\,(T/M_p)^3$, so gravitons could {\it
%not} be kept in equilibrium with the surrounding plasma for
%$T\,<\,M_p$. Therefore, GW produced well after Planck scale will
%always be decoupled from the plasma, 
and, whatever their spectral
signatures, they will retain their shape throughout the expansion of
the Universe. Thus, the characteristic frequency and shape of the GWB
generated at a given time should contain information about the very
early state of the Universe in which it was produced. Actually, it
is conceivable that, in the not so far future, the detection of these
GW backgrounds could be the only way we may have to infer the physical
conditions of the Universe at such high energy scales. % (which certainly
%no particle collider will ever reach). 
However, the same reason that
makes GW ideal probes of the early universe $-$ the weakness of gravity
$-$ is responsible for the extreme difficulties we have for their
detection on Earth. %For an extensive discussion see Ref.~\cite{Thorne}.

In %a recent letter
Refs.~\cite{GBF,GBFS} we described the stochastic background
predicted to arise from reheating after hybrid inflation. Here
%In this paper 
we will review the various processes involved in
the production of such a background.
%, whose detection could open a new
%window into the very early universe. 
In the future, this background
could serve as a new tool to discriminate among different
inflationary models, since reheating in each model would give 
rise to a different GWB with very characteristic spectral features.
% The first stage of the
%energy conversion at the end of inflation,
%preheating~\cite{preheating}, is known to be explosive and extremely
%violent, and quite often generates in less than a Hubble time the huge
%entropy measured today. 
The details of the dynamics of preheating
depend very much on the model and are often very complicated because
of the non-linear, non-perturbative and out-of-equilibrium character
of the process itself. However, all the cases have in common that only
specific resonance bands of the fields suffer an exponential
instability, which makes their occupation numbers grow by many orders
of magnitude. The shape and size of the spectral bands depend very
much on the inflationary model. If one translates this picture into
position-space, the highly populated modes correspond to large
time-dependent inhomogeneities in the matter distributions which
acts, in fact, as a powerful source of GW.
%the source of GW we are looking for.
For example, in single field chaotic inflation models, the coherent
oscillations of the inflaton during preheating generates, via
parametric resonance, a population of highly occupied modes that
behave like waves of matter. They collide among themselves and their
scattering leads to homogenization and local thermal equilibrium.
These collisions occur in a highly relativistic and very asymmetric
way, being responsible for the generation of a stochastic
GWB~\cite{TkachevGW,JuanGW,EastherLim,EastherLim2,DufauxGW} with a typical frequency today
of the order of $10^{7} - 10^{9}$ Hz, corresponding to the present
size of the causal horizon at the end of high-scale inflation. 
%There
%is at present a couple of experiments searching for such a background,
%see Refs.~\cite{Nishizawa2007}, based of laser interferometry, as well
%as by resonant superconducting microwave cavities~\cite{Picasso}.

However, there are models like hybrid inflation in which the end of
inflation is sudden~\cite{hybrid} and the conversion into radiation
occurs almost instantaneously. Indeed, %since the work of
%Ref.~\cite{tachyonic} we know that 
hybrid models preheat very violently, 
%more violent way than chaotic inflation models, 
via the spinodal instability of the symmetry breaking field 
that triggers the end of
inflation, irrespective of the couplings that this field may have to
the rest of matter. Such a process is known as \textit{tachyonic
preheating}~\cite{tachyonic,symmbreak} and could be responsible for
copious production of dark matter particles~\cite{ester}, lepto and
baryogenesis~\cite{CEWB}, topological defects~\cite{tachyonic},
primordial magnetic fields~\cite{magnetic}, etc.
%It was speculated in Ref.~\cite{JuanGW} that in (low-scale) models of
%hybrid inflation it might be possible to generate a stochastic GWB in
%the frequency range accessible to present detectors, if the scale of
%inflation was as low as $H_{\rm inf} \sim 1$ TeV. However, the
%amplitude was estimated using the parametric resonance formalism of
%chaotic preheating, which may not be applicable in this case. 
In Ref.~\cite{symmbreak}, it was
shown that the process of symmetry breaking in hybrid preheating, 
proceeds via the
nucleation of dense bubble-like structures moving at relativistic
speeds, which collide and break up into smaller structures (see Figs.~7
and~8 of Ref.~\cite{symmbreak}). We conjectured at that time that such collisions
would be a very strong source of GW, analogous to the GW production
associated with strongly first order phase
transitions~\cite{KosowskyTurner}. As we will show here, this
is indeed the case during the nucleation, collision and subsequent
rescattering of the initial bubble-like structures produced after
hybrid inflation. During the different stages of reheating in this
model, gravity waves are generated and amplified until the Universe
finally thermalizes and enters into the radiation era of the
Standard Model of Cosmology. From that moment until now, %during the
%whole thermal history of the expansion of the universe, 
this cosmic GWB will be redshifted as a radiation-like fluid, 
totally decoupled from any other energy-matter content of the universe, 
such that today's ratio of energy stored in these GW to that in radiation, 
could range from $\Omega_{_{\rm GW}}h^2\sim 10^{-8}$, peacked around $f\sim
10^7$~Hz for the high-scale models, to $\Omega_{_{\rm GW}}h^2\sim
10^{-11}$, peacked around $f\sim 1$ Hz for the low-scale models.

Finally, since the first paper by Khlebnikov and
Tkachev~\cite{TkachevGW}, studing the GWB produced at reheating after
chaotic inflation, there has been some developments. 
The idea was soon extended to hybrid inflation in Ref.~\cite{JuanGW}. It was
also revisited very recently in Ref.~\cite{EastherLim,EastherLim2} for
the $\lambda\phi^4$ and $m^2\phi^2$ chaotic scenarios, and reanalysed
again for hybrid inflation in Refs.~\cite{GBF,GBFS}, using the new
formalism of tachyonic preheating~\cite{tachyonic,symmbreak}. Because
of the increase in computer power of the last few years, we are now
able to perform precise simulations of the reheating process in a
reasonable time scale. Moreover, understanding of reheating has
improved, while gravitational waves detectors are beginning to attain
the aimed sensitivity~\cite{LIGO}. Furthermore, since these cosmic
GWBs could serve as a deep probe into the very early universe, we
should characterize in the most detailed way the information that we
will be able to extract from them.

\section{Gravitational Wave Production}

Our main purpose here is to study the details of the
stochastic GWB produced during the reheating stage after
hybrid inflation (sections 2 and 3). Nevertheless, 
%However, 
we also study more briefly
%albeit very briefly, 
the analogous background from reheating in some
chaotic models (section 4). Thus, in this section we derive a
general formalism for extracting the GW power spectrum in any scenario
of reheating within the (flat) Friedman-Robertson-Walker (FRW)
universe. The formalism will be simplified when applied to scenarios
in which we can neglect the expansion of the universe, like in the
case of Hybrid models.

A theory with an inflaton scalar field $\chi$ interacting 
with other Bose fields $\phi_a$, can be described by
\begin{equation}\label{lagrangian}
\mathcal{L} = \frac{1}{2}\partial_\mu\chi\partial^\mu\chi +
\frac{1}{2}\partial_\mu\phi_a\partial^\mu\phi_a + \frac{R}{16\pi G} -
V(\phi,\chi)\,
\end{equation}
with $R$ the Ricci scalar. For hybrid models, we consider a generic
symmetry breaking `Higgs' field $\Phi$, with $N_c$ real components. 
We can take $\Phi^{\dagger}\Phi = {1\over2}\sum_a\phi_a^2 \equiv
|\phi|^2/2$, with $a$ running for the number of Higgs' components, 
\textit{e.g.} $N_c = 1$ for a real scalar Higgs,
$N_c = 2$ for a complex scalar Higgs or $N_c = 4$ for a $SU(2)$ Higgs,
etc. The effective potential then becomes
\begin{equation}\label{higgsPotentialII}
V(\phi,\chi) = \frac{\lambda}{4}\left(|\phi|^2 - 
v^2\right)^2 + g^2\chi^2|\phi|^2
+ \frac{1}{2}\mu^2\chi^2\,.
\end{equation}
For chaotic scenarios, we consider a massless scalar field $\phi$
interacting with the inflaton $\chi$ via
\begin{equation}
V(\chi,\phi) = \frac{1}{2}g^2\chi^2\phi^2 + V(\chi)\,,
\end{equation}
with $V(\chi)$ the inflaton's potential. Concerning the 
simulations we show in this paper, we concentrate in the 
$N_c = 4$ case for the hybrid model and consider a potential 
$V(\chi) = {\lambda\over4}\chi^4$ for the chaotic scenario.

The classical equations of motion of the
inflaton and the other Bose fields are
\begin{eqnarray}
\label{inflatonEq}
\ddot\chi + 3H\dot\chi - \frac{1}{a^2}\nabla^2\chi + \frac{\partial
V}{\partial\chi} = 0\,, \hspace{1cm}
%\label{scalarEq}
%&&
\ddot\phi_a + 3H\dot\phi_a - \frac{1}{a^2}\nabla^2\phi_a +
\frac{\partial V}{\partial\phi_a} = 0
\end{eqnarray}
with $H = \dot a/a$. On the other hand, GW are represented here 
by a transverse-traceless (TT) gauge-invariant metric perturbation,
$h_{ij}$, on top of the flat FRW space 
$ds^2 = -dt^2 + a^2(t)\left(\delta_{ij} + h_{ij}\right)dx^idx^j$,
with $a(t)$ the scale factor and the tensor perturbations verifying 
$\partial_ih_{ij} = h_{ii} = 0$. % (In the following, we will raise 
%or low indices of the metric perturbations with the delta Kronecker 
%$\delta_{ij}$, so $h_{ij} = h^{i}_{j} = h^{ij}$ and so on). 
Then, the Einstein field equations 
can be splitted into the
background %$G_{\mu\nu}^{(0)} = 8\pi G\,T_{\mu\nu}^{(0)}$ 
and the perturbed %$\delta G_{\mu\nu} = 8\pi G\,\delta {\rm T}_{\mu\nu}$
equations. The former describe the evolution of the flat
FRW universe through
\begin{eqnarray}
\label{hubbleDotEq}
-\frac{\dot H}{4\pi G} &=&  \dot\chi^2 +
\frac{1}{3a^2}(\nabla\chi)^2 + \dot\phi_a^2 +
\frac{1}{3a^2}(\nabla\phi_a)^2\\
\label{hubbleEq}
\frac{3H^2}{4\pi G} &=& \dot\chi^2 +
\frac{1}{a^2}(\nabla\chi)^2 + \dot\phi_a^2 +
\frac{1}{a^2}(\nabla\phi_a)^2 + 2V(\chi,\phi)
\end{eqnarray}
where any term in the r.h.s. of~(\ref{hubbleDotEq}) and~(\ref{hubbleEq}), 
should be understood as spatially averaged.

On the other hand, the perturbed Einstein equations describe the evolution 
of the tensor perturbations~\cite{Mukhanov} as
\begin{equation}\label{GWeq}
 \ddot h_{ij} + 3H\dot h_{ij} - \frac{1}{a^2}\nabla^2h_{ij} =
16\pi G\,\Pi_{ij}\,,
\end{equation}
with $\partial_i\Pi_{ij} = \Pi_{ii} = 0$. 
The source of the GW, $\Pi_{ij}$, 
contributed by both the inflaton and the other scalar fields, 
will be just the transverse-traceless part of the (spatial-spatial) 
components of the total anisotropic stress-tensor
\begin{eqnarray}\label{ast}
{\rm T}_{\mu\nu} = \left[ \partial_\mu\chi
\partial_\nu\chi + \partial_\mu\phi_a\partial_\nu\phi_a + 
g_{\mu\nu}(\mathcal{L} - \left\langle p \right\rangle)\right]/a^2,
\end{eqnarray}
where $\mathcal{L}(\chi,\phi_a)$ is the lagrangian~(\ref{lagrangian})
and $\left\langle p \right\rangle$ is the background homogeneous
pressure. As we will explain in the next subsection, when extracting
the TT part of~(\ref{ast}), the term proportional to $g_{\mu\nu}$ in
the r.h.s of~(\ref{ast}), will be dropped out from the GW equations
of motion. Thus, the effective source of the GW will be just
given by the TT part of the gradient terms
$\partial_\mu\chi\partial_\nu\chi +
\partial_\mu\phi_a\partial_\nu\phi_a$.

%In summary, Eqs.~(\ref{inflatonEq})-(\ref{scalarEq}), together with
%Eqs.~(\ref{hubbleDotEq})-(\ref{hubbleEq}), describe the coupled
%dynamics of reheating in any inflationary scenario, while
%Eq.~(\ref{GWeq}) describe the production of GW in each of those
%scenarios. 
%In this paper we use lattice simulations to study the
%generation of GW during reheating. %after inflation.
%Specific details on this are given in section~IV, 
%but let us just mention here that 
%Our
%approach is to solve the evolution of the gravitational waves
%simultaneously with the dynamics of the scalar fields, in a
%discretized lattice with periodic boundary conditions. We assume
%initial quantum fluctuations for all fields and only a zero mode for
%the inflaton. Moreover, we also included the GW backreaction on the
%scalar fields' evolution via the gradient terms,
%$h^{ij}\nabla_i\chi\nabla_j\chi + h^{ij}\nabla_i\phi_a\nabla_j\phi_a$
%and confirmed that, for all practical purposes, these are negligible
%throughout GW production.

\subsection{The Transverse-Traceless Gauge}

A generic (spatial-spatial) metric perturbation $\delta h_{ij}$ has six 
independent degrees of freedom, whose contributions can be split 
into~\cite{Mukhanov} scalar, vector and tensor metric perturbations 
%\begin{equation}
%\indent
$\delta h_{ij} = \psi\,\delta_{ij} + E_{,ij} + F_{(i,j)} + h_{ij}\,$, 
%\end{equation}
with $\partial_iF_i = 0$ and $\partial_ih_{ij} = h_{ii} = 0$.  By
choosing a transverse-traceless stress-tensor source $\Pi_{ij}$, we
can eliminate all the degrees of freedom (d.o.f.) but the pure TT
part, $h_{ij}$, which represent the only physical d.o.f which
propagate and carry energy out of the source (i.e. GW). %If we had chosen
%only a traceless but non-transverse stress source, we could have
%eliminated the scalar d.o.f.  $\psi$ and absorbed $E$ into the scalar
%field perturbation, but we would still be left with a vector field
%$F_i$ also sourced by the (traceless but non-transverse) anisotropic
%stress tensor, thus giving rise to a vorticity divergence-less field
%$V_i$. However, since the initial conditions are those of a scalar
%Gaussian random field (see section IV), even in that case of a
%non-transverse but traceless stress source, the mean vorticity of the
%subsequent matter distribution, averaged over a sufficiently large
%volume, should be zero (although locally we do have vortices of the
%Higgs field, see Refs.~\cite{CEWB,magnetic}), since vortices with
%opposite chirality cancel eachother. This means that in such a case,
%although $\partial^i\Pi_{ij} \neq 0$, and thus $\partial^i\delta
%h_{ij} \neq 0$, we could still recover the TT component when averaging
%over the whole realization.
Thus, taking the TT part of the anisotropic stress-tensor, 
we ensure that we only source the physical d.o.f. 
that represent GW.
%For practical purposes, we will 
%Thus, 
%considering from the beggining 
%The equations of motion
%of the TT metric perturbations are then given by Eq.~(\ref{GWeq}).
%Note that for a non-transverse source the equations would have been
%much more complicated, so the advantage of using the TT part from the
%beginning is clear. The disadvantage arises because 
%Unfortunatelly, obtaining the TT
%part of a tensor %in configuration space 
%is very demanding in 
%computational time. However, as we will explain next, 
%we use a method by which we can circunvent 
%this issue. Let's see it.

Let us switch to Fourier space. %Using the convention
%\begin{equation}
%\indent f(\mathbf{k}) = \frac{1}{(2\pi)^{3/2}}\int d^3\mathbf{x}\,
%e^{+i\mathbf{kx}}f(\mathbf{x})\,,
%\end{equation} 
The GW equations~(\ref{GWeq}) then %in Fourier space
read
\begin{equation}\label{GWeqFourier}
 \ddot h_{ij}(t,\mathbf{k}) + 3H\dot h_{ij}(t,\mathbf{k}) + 
\frac{k^2}{a^2}h_{ij}(t,\mathbf{k}) = 16\pi G\,\Pi_{ij}(t,\mathbf{k})\,,
\end{equation}
where $k = |\mathbf{k}|$. Assuming no GW at the beginnig of reheating
(i.e.  the end of inflation $t_e$), the initial conditions are
$h_{ij}(t_e) = \dot h_{ij}(t_e) = 0$, so the solution to
Eq.~(\ref{GWeqFourier}) for $t > t_e$ will be just given by a causal
convolution with an appropriate Green's function $G(t,t')$,
\begin{eqnarray}\label{GWsol}
 h_{ij}(t,\mathbf{k}) = 16\pi G \int_{t_e}^{t}dt'\,G(t,t')
\Pi_{ij}(t',\mathbf{k})\,.
\end{eqnarray} 
Therefore, all we need to know for computing the GW is the TT part of
the stress-tensor, $\Pi_{ij}$, and the Green's function $G(t',t)$.
However, as we will demonstrate shortly, we have used a numerical
method by which we don't even need to know the actual form of
$G(t',t)$. To see this, let us extract the TT part of the total
stress-tensor. Given the symmetric anisotropic stress-tensor
T$_{\mu\nu}$~(\ref{ast}), we can easily obtain the TT part of its
spatial components in momentum space, $\Pi_{ij}(\mathbf{k})$. Using
the spatial projection operators $P_{ij} = \delta_{ij} - \hat k_i \hat
k_j$, with $\hat k_i = k_i/k$, then~\cite{Carroll} 
$\Pi_{ij}(\mathbf{k}) = \Lambda_{ij,lm}(\mathbf{\hat k})
{\rm T}_{lm}(\mathbf{k})$,
where
\begin{eqnarray}\label{projector}
 \Lambda_{ij,lm}(\mathbf{\hat k}) \equiv \Big(P_{il}(\mathbf{\hat k})
P_{jm}(\mathbf{\hat k}) - {1\over2} P_{ij}(\mathbf{\hat k})
P_{lm}(\mathbf{\hat k})\Big)\,.
\end{eqnarray}
%is the projection operator.
Thus, one can easily see that, at any time $t$, $k_i\Pi_{ij}(\mathbf{\hat
k},t) = \Pi_{i}^{i}(\mathbf{\hat k},t) = 0$, as required, thanks to
the identities $P_{ij}\hat k_j = 0$ and $P_{ij}P_{jm} = P_{im}$.

Note that %the TT tensor, $\Pi_{ij}$, is just a linear combination of
%the components of non-traceless nor-transverse tensor
%T$_{ij}$~(\ref{ast}), while 
the solution~(\ref{GWsol}) is just linear
of the non-traceless nor-transverse tensor
T$_{ij}$~(\ref{ast}).
%$\Pi_{ij}$.
Therefore, we can write the TT tensor perturbations 
(i.e. the GW) as
\begin{eqnarray}\label{TTsol}
h_{ij}(t,\mathbf{k}) = \Lambda_{ij,lm}(\mathbf{\hat k}) 
u_{lm}(t,\mathbf{k}),
\end{eqnarray} 
with $u_{ij}(t,\mathbf{k})$ the Fourier transform of the solution 
of the following equation
\begin{eqnarray}\label{GWfakeEq}
 \ddot u_{ij} + 3H\dot u_{ij} - 
\frac{1}{a^2}\nabla^2 u_{ij} = 16\pi G\,{\rm T}_{ij}\,.
\end{eqnarray}
This Eq.~(\ref{GWfakeEq}) is nothing but Eq.~(\ref{GWeq}), sourced
with the complete T$_{ij}$~(\ref{ast}), instead of with its TT part,
$\Pi_{ij}$. Of course, Eq.~(\ref{GWfakeEq}) contains unphysical
(gauge) d.o.f.; however, in order to obtain the real physical TT
d.o.f. $h_{ij}$, we can evolve Eq.(\ref{GWfakeEq}) in configuration
space, Fourier transform its solution and apply the
projector~(\ref{projector}) as in ~(\ref{TTsol}). This way we can
obtain in momentum space, at any moment of the evolution, the physical
TT d.o.f. that represent GW, $h_{ij}$. Whenever needed, we can Fourier
transform back to configuration space and obtain the spatial
distribution of the gravitational waves.

Moreover, since the second term of the r.h.s of the total
stress-tensor T$_{ij}$ is proportional to $g_{ij} = \delta_{ij} +
h_{ij}$, see~(\ref{ast}), when aplying the TT
projector~(\ref{projector}), the part with the $\delta_{ij}$ just
drops out, simply because it is a pure trace, while the other part 
contributes with a term $-(\mathcal{L} - \left\langle p
\right\rangle)h_{ij}$ in the l.h.s of Eq.(\ref{GWeqFourier}). 
%Since
%$(\mathcal{L} - \left\langle p \right\rangle) = - (16\pi
%G/3)\left\langle\partial_i\chi\partial^i\chi + \partial_i\phi^a
%\partial^i\phi_a\right\rangle$,
However, $(\mathcal{L} - \left\langle p \right\rangle)$ is of the same 
order as the metric perturbation $\sim {\cal O}(h)$, 
so this extra term is second order in the gravitational coupling 
and it can be neglected in the GW Eqs.~(\ref{GWeqFourier}).
This way, the effective source in Eq.~(\ref{GWfakeEq}) is just the 
gradient terms of both the inflaton and the other scalar fields,
\begin{eqnarray}\label{GWfakeSource}
{\rm T}_{ij} = (\nabla_i\chi\nabla_j\chi + 
\nabla_i\phi_a\nabla_j\phi_a)/a^2.
\end{eqnarray} 
Therefore, the effective source of the physical GW, will be just the
TT part of~(\ref{GWfakeSource}), as we had already mentioned before.

%Alternatively, one could evolve the equation of the TT tensor
%perturbation in configuration space, Eq.~(\ref{GWeq}), with the source
%given by
%\begin{equation}\label{GWtrueSource}
%\Pi_{ij}({\bf x},t) = {1\over(2\pi)^{3/2}}\int d^3{\bf k}\,
%e^{-i{\bf kx}}\Lambda_{ij,lm}({\bf \hat k}){\rm T}_{lm}({\bf k},t)\,, 
%\end{equation}
%such that $\partial_i\Pi_{ij}({\bf x},t) = \Pi_{ii}({\bf x},t) = 0$ at
%any time.  So, at each time step of the evolution of the fields, one
%would have first to compute (the gradient part of)
%T$_{lm}$~(\ref{GWfakeSource}) in configuration space, then Fourier
%transform it to momentum space, substitute in Eq.~(\ref{GWtrueSource})
%and perform the integral. However, proceeding as we suggested above,
%there is no need to perform the integral,
%\footnote{Of course, if we want to obtain the spatial distribution
%of the tensor perturbations, we have to perform an inverse Fourier
%transform from $h_{ij}({\bf k},t)$ to $h_{ij}({\bf x},t)$. However, we
%can do this only at those times when we are interested on taking a
%snapshot of the GW spatial distribution, but not at every time step of
%the evolution of the fields.}, 
%nor Fourier transform the fields at each time step, but rather only at
%those times at which we want to measure the GW spectrum. The viability
%of our method relies in the following observation. To compute the GW
%we could, first of all, project the TT part of the
%source~(\ref{GWtrueSource}), and second, solve Eq.~(\ref{GWeq}).
%However, we achieve the same result if we commute these two operations
%such that,

We have found the \textit{commuting procedure} proposed (i.e. the fact 
that we first solve Eq.~(\ref{GWfakeEq}) and secondly we apply the TT 
projector to the solution~(\ref{TTsol}), and not the other way around), 
very useful. We are able to extract the spectra or the spatial 
distribution of the GW at any desired time, saving a great amount of 
computing time since we don't have to be Fourier transforming the 
source at each time step. Most importantly, with this procedure we can 
take into account backreaction simultaneously with the fields evolution.

In summary, for solving the dynamics of reheating of a particular
inflationary model, we evolve Eqs.~(\ref{inflatonEq}) %-(\ref{scalarEq})
in the lattice, together with
Eqs.~(\ref{hubbleDotEq})-(\ref{hubbleEq}), while for the GWs 
%, instead of solving Eq.~(\ref{GWeq}), 
we solve Eq.~(\ref{GWfakeEq}). Then, only
when required, we Fourier transform the solution of
Eq.~(\ref{GWfakeEq}) and then apply~(\ref{TTsol}) in order to recover
the physical transverse-traceless d.o.f representing the GW. From
there, one can easily build the GW spectra or take a snapshot of
spatial distribution of the gravitational waves.

\subsection{The energy density in GW}

The energy-momentum tensor of the GW is given by~\cite{Carroll}
\begin{equation}\label{tmunu}
 t_{\mu\nu} = {1\over 32\pi G}\,\left\langle\partial_\mu
h_{ij}\, \partial_\nu h^{ij}\right\rangle_{\rm V}\,,
\end{equation}
where $h_{ij}$ are the TT tensor perturbations solution of
Eq.~(\ref{GWeq}). The expectation value
$\left\langle...\right\rangle_{\rm V}$ is taken over a region of
sufficiently large volume $V=L^3$ to encompass enough physical
curvature to have a gauge-invariant measure of the GW energy-momentum
tensor.

The GW energy density will be just 
$\rho_{_{\rm GW}} = t_{00}$, so
\begin{eqnarray}\label{rho}
 \rho_{_{\rm GW}} = \frac{1}{32\pi G}\frac{1}{L^3}\int d^3\mathbf{x}\,
\dot h_{ij}(t,\mathbf{x})\dot h_{ij}(t,\mathbf{x}) =  
\frac{1}{32\pi G}\frac{1}{L^3}\int d^3\mathbf{k}\,
\dot h_{ij}(t,\mathbf{k})\dot h_{ij}^*(t,\mathbf{k})\,,
\end{eqnarray} 
where in the last step we Fourier transformed each 
$\dot h_{ij}$ and used the %integral 
definition of the Dirac delta.
% $(2\pi)^3\delta^{(3)}(\mathbf{k}) = \int d^3\mathbf{x}\ e^{-i\mathbf{kx}}$.
We can always write the scalar product in (\ref{rho}) in terms of the 
(Fourier transformed) solution $u_{lm}$ of the Eq.(\ref{GWfakeEq}), 
by just using %that the
%spatial projectors (\ref{projector})
%\begin{equation}
% \dot h_{ij}\dot h_{ij} = \Lambda_{ij,lm}\Lambda_{ij,rs} 
%\dot u_{lm} \dot u_{rs} = 
%\Lambda_{lm,rs} \dot u_{lm} \dot u_{rs}\,,
%\end{equation}
%where we have used 
the fact that $\Lambda_{ij,lm}\Lambda_{lm,rs} =
\Lambda_{ij,rs}$. This way, we can express the GW energy density as
\begin{eqnarray}\label{rhoTotal}
\rho_{_{\rm GW}} = \frac{1}{32\pi G L^3}
\int k^2dk \int d\Omega\,\Lambda_{ij,lm}(\mathbf{\hat k})
\dot  u_{ij}(t,\mathbf{k})\dot  u_{lm}^*(t,\mathbf{k}).
\end{eqnarray}
From here, we can also compute the power spectrum per logarithmic frequency
interval in GW, normalized to the critical density $\rho_c$, as 
$\Omega_{_{\rm GW}}=\int {df\over f}\,\Omega_{_{\rm GW}}(f)\,,$ where
\begin{eqnarray}\label{OmegaFraction}
\Omega_{_{\rm GW}}(k) \equiv \frac{1}{\rho_c}\frac{d\rho_{_{\rm GW}}}{d\,{\rm log}k}
= \frac{k^3}{32\pi GL^3\rho_c} \int d\Omega\,\Lambda_{ij,lm}(\mathbf{\hat k})
\dot  u_{ij}(t,\mathbf{k})\dot  u_{lm}^*(t,\mathbf{k})\hspace{.4cm} 
\end{eqnarray}
We have checked explicitely in the simulations that the argument 
of the angular integral of~(\ref{OmegaFraction}) is independent 
of the directions in {\bf k}-space. Thus, whenever we plot the GW 
spectrum of any model, we will be showing the amplitude of the 
spectrum (per each mode $k$) as obtained after avaraging over all 
the directions in momentum space,
\begin{eqnarray}
\Omega_{_{\rm GW}}(k) = \frac{k^3}{8GL^3\rho_c} \left\langle \Lambda_{ij,lm}(\mathbf{\hat k})
\dot  u_{ij}(t,\mathbf{k})\dot  u_{lm}^*(t,\mathbf{k})\right\rangle_{4\pi} 
\end{eqnarray} 
with $\left\langle f \right\rangle_{4\pi} \equiv {1\over 4\pi}\int f{\rm d}\Omega$. 

Finally, we must address the fact that the frequency range, for a GWB
produced in the early universe, will be redshifted today. We should
calculate the characteristic physical wavenumber of the present GW
spectrum, which is redshifted from any time $t$ during GW
production. %This is a key point, since a relatively long period of
%turbulence will develop after preheating, which could change the
%amplitude of the GWB and shift the frequency at which the spectra
%peaks. 
So let us distinguish four characteristic times: the end of
inflation, $t_e$; the time $t_*$ when GW production stops;
the time $t_{r}$ when the universe finally reheats and enters into the
radiation era; and today, $t_0$. %\footnote{Note, however, that after
%thermalization there is still a small production of GW from the
%thermal plasma, but this can be ignored for all practical purposes.}
Thus, today's frequency $f_0$ is related to the physical wavenumber
$k_t$ at any time $t$ of GW production, via $f_0 =
(a_t/a_0)(k_t/2\pi)$, with $a_0$ and $a_t$, the scale factor today and
at the time $t$, respectively. Thermal equilibrium was established at
some temperature $T_r$, at time $t_r \geq t$. The Hubble rate at that
time was $M_P^2H_r^2 = (8\pi/3)\rho_r$, with $\rho_r =
g_r\pi^2T_r^4/30$ the relativistic energy density and $g_r$ the
effective number of relativistic degrees of freedom at temperature
$T_r$. Since then, the scale factor has increased as $a_r/a_0 =
(g_{0,s}/g_{r,s})^{1/3}(T_0/T_{r})$, with $g_{i,s}$ the effective
entropic degrees of freedom at time $t_i$, and $T_0$
today's CMB temperature. Putting all together,
\begin{eqnarray}
\label{redshift}
 f_0 = \left(\frac{8\pi^3g_{r}}{90}\right)^{1\over4}
\left(\frac{g_{0,s}}{g_{{r},s}}\right)^{1\over3}\frac{T_0}
{\sqrt{H_{r}M_p}}\left(\frac{a_e}{a_r}\right)\frac{k}{2\pi}\,,
\end{eqnarray}
where we have used the fact that the physical wave number $k_t$ at 
any time $t$ during GW production, is related to the comoving wavenumber $k$
through $k_t = (a_e/a_t)k$ with the normalization $a_e \equiv 1$.

From now on, we will be concerned with hybrid inflation, leaving
chaotic inflation for section 4. Within the hybrid scenario, we will
analyse the dependence of the shape and amplitude of the produced GWB
on the scale of hybrid inflation, and more specifically on the
\textit{v.e.v.} of the Higgs field triggering the end of inflation.
%The initial conditions are carefully treated following the
%prescription adopted in paper I, as explained in section IV. 
Given the natural frequency at hand in hybrid models, $m = \sqrt{\lambda}v$,
whose inverse $m^{-1}$ sets the characteristic time scale during the
first stages of reheating, it happens that as long as $v\,\ll\,M_p$,
the Hubble rate $H\,\sim\,\sqrt{\lambda}(v^2/M_p)$ is much smaller
than such a frequency, $H\,\ll\,m$. Indeed, all the initial vacuum
energy $\rho_0$ gets typically converted into radiation in less than a
Hubble time, in just a few $m^{-1}$ time steps. Therefore, we should
be able to ignore the dilution due to the expansion of the universe
during the production of GW, at least during the first stages of
reheating. %However, as we will see, the turbulent behaviour developed
%after those first stages, could last for much longer than an e-fold,
%in which case we will have to take into account the expansion of the
%universe. 
Our approach will be to ignore the expansion of the Universe, such 
% and later see how we can account for corrections if needed. Thus, 
that we fix the scale factor to one, $a = 1$. % and the Hubble rate $H =
%0$ and $\dot H = 0$. 
As we will see later, neglecting the expansion of 
the Universe for the time of GW production, 
will be completely justified
\textit{a~posteriori}.

The system of equations that
we have to solve numerically in a lattice for the hybrid model are
\begin{eqnarray}\label{GWeqHybrid}
&&\ddot u_{ij} - \nabla^2 u_{ij} = 16\pi G\,{\rm T}_{ij} \\
&&\ddot\chi - \nabla^2\chi + \left(g^2|\phi|^2 +
\mu^2\right)\chi = 0 \\ &&\ddot\phi_a - \nabla^2\phi_a +
\left(g^2\chi^2 + \lambda|\phi|^2 - m^2\right)\phi_a = 0
\end{eqnarray}
with T$_{ij}$ given by Eq.(\ref{GWfakeSource}) with the scale factor
$a = 1$. We have explicitly checked in our computer simulations that
the backreaction of the gravity waves into the dynamics of both the
inflaton and the Higgs fields is negligible and can be safely ignored.
We thus omit the backreaction terms in the above equations.

We evaluated during the evolution of the system the mean field values,
as well as the different energy components. Initially, %As shown in
%Fig.~\ref{fig1}, 
the Higgs field grows towards the true vacuum and
the inflaton moves towards the minimum of its potential and oscillates
around it. We have checked that the sum of the averaged gradient,
kinetic and potential energies (contributed by both the inflaton and
the Higgs), remains constant during reheating, as expected, since the
expansion of the universe is irrelevant in this model. We have also
checked that the time evolution of the different energy components is
the same for different lattices, changing the number of points
$N$, the minimum momentum $p_{\rm min} = 2\pi/L$ or
the lattice spacing $a = L/N$, with $L$ the lattice size. %, as long
%as the product $ma < 0.5$; for a detailed discussion of lattice scales
%see paper I.  
The evolution of the Higgs' \textit{v.e.v.} 
%in  Fig.~\ref{fig1}, 
follows three stages easily distinguished. 
First, an exponential growth of the \textit{v.e.v.}
towards the true vacuum. This is driven by the tachyonic instability
of the long-wave modes of the Higgs field, that makes the spatial
distribution of this field to form lumps and bubble-like
structures~\cite{tachyonic,symmbreak}. Second, the Higgs field
oscillates around the true vacuum, as the Higgs' bubbles collide and
scatter off eachother. Third, a period of turbulence is reached,
during which the inflaton oscillates around its minimum and the Higgs
sits in the true vacuum. For a detailed description of the dynamics of
these fields see Ref.~\cite{symmbreak}. Here we will be only concerned
with the details of the gravitational wave production.

%%%%%%%%%%%%%%%%%%%%%%%%%
\begin{figure}[t]
%\vspace{5mm}
\begin{center}
\includegraphics[width=10cm,angle=-90]{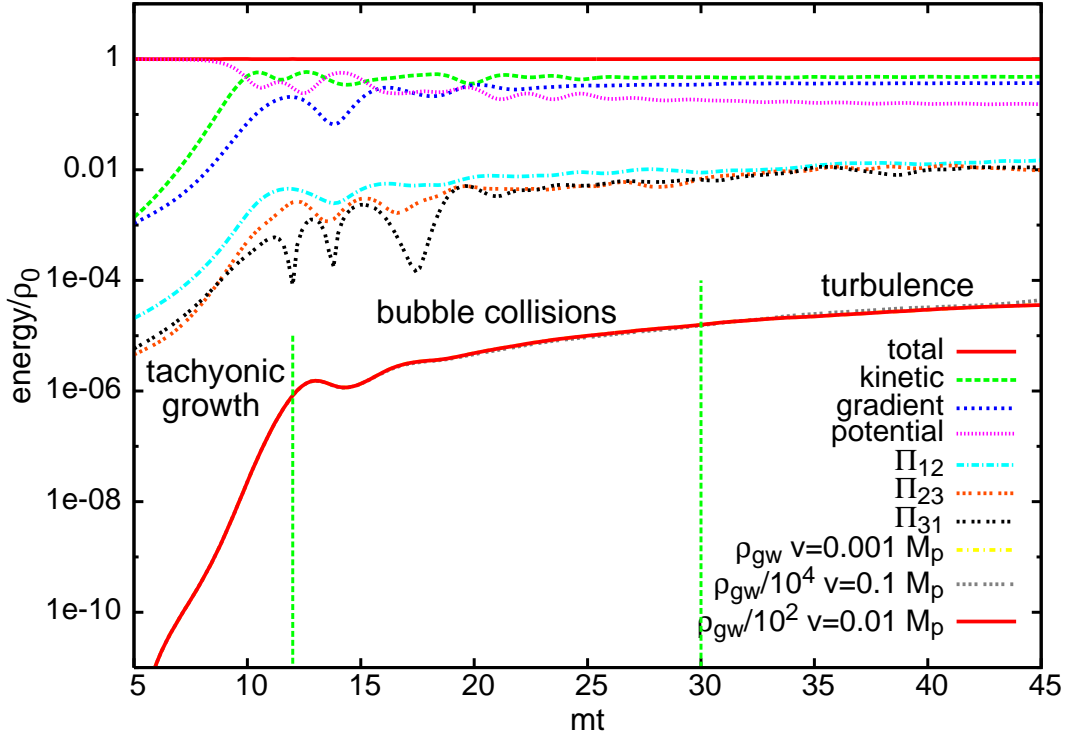}
\end{center}
\vspace*{-5mm}
\caption{The time evolution of the different types of energy (kinetic,
gradient, potential, anisotropic components and gravitational waves
for different lattices), normalized to the initial vacuum energy,
after hybrid inflation, for a model with $v=10^{-3}\,M_P$. One can
clearly distinguish here three stages: tachyonic growth, bubble
collisions and turbulence. }
\label{fig3}
\vspace*{-3mm}
\end{figure}
%%%%%%%%%%%%%%%%%%%%%%%%%

The initial energy density at the end of hybrid inflation is given by
$\rho_0 = m^2v^2/4$, with $m^2 = \lambda v^2$, so the fractional
energy density in gravitational waves is
\begin{equation}
 {\rho_{_{\rm GW}}\over\rho_0} = {4t_{00}\over v^2\,m^2} =
\frac{1}{8\pi\,G v^2 m^2}\left\langle\dot h_{ij}\dot
h^{ij}\right\rangle_{\rm V}\,,
\end{equation}
where $\left\langle\dot h_{ij}\dot h^{ij}\right\rangle_V$, defined as a 
volume average like 
${1\over {\rm V}}\int d^3{\bf x}\dot h_{ij}\dot h^{ij}$, is extracted from 
the simulations as
\begin{eqnarray}
\left\langle\dot h_{ij}\dot
h^{ij}\right\rangle_{\rm V} = \frac{4\pi}{V}\int d{\rm log}
k\,k^3\left\langle \Lambda_{ij,lm}(\mathbf{\hat k})
\dot  u_{ij}(t,\mathbf{k})\dot  u_{lm}^*(t,\mathbf{k})
\right\rangle_{4\pi} 
\end{eqnarray}
where $u_{ij}(t,\mathbf{k})$ is the Fourier transform of the solution 
of Eq.~(\ref{GWeqHybrid}). 
Then, we can compute the corresponding
density parameter today (with $\Omega_{\rm
rad}\,h^2\simeq3.5\times10^{-5}$)
\begin{eqnarray}
\Omega_{_{\rm GW}}\,h^2 = {\Omega_{\rm rad}\,h^2\over 2
G\,v^2\,m^2\,V}\,\int d{\rm log}k\, k^3\left\langle
\Lambda_{ij,lm}(\mathbf{\hat k})
\dot  u_{ij}(t,\mathbf{k})\dot  u_{lm}^*(t,\mathbf{k})
\right\rangle_{4\pi} %\nonumber \\
\end{eqnarray}
which has assumed that all the
vacuum energy $\rho_0$ gets converted into radiation, an approximation
which is always valid in generic hybrid inflation models with $v\ll
M_P$, and thus $H\ll m=\sqrt\lambda\,v$.

We have shown in Fig.~\ref{fig3} the evolution in time of the fraction
of energy density in GW. The first (tachyonic) stage is clearly
visible, with a (logarithmic) slope twice that of the anisotropic
tensor $\Pi_{ij}$. Then there is a small plateau corresponding to the
production of GW from bubble collisions; and finally there is the
slow growth due to turbulence. In the next section we will describe
in detail the most significant features appearing at each stage.

Note that in the case that $H\ll m$, the maximal production of GW
occurs in less than a Hubble time, soon after symmetry breaking, while
turbulence lasts several decades in time units of $m^{-1}$. Therefore,
we can safely ignore the dilution due to the Hubble expansion, up to
times much greater than those of the tachyonic instability. Eventually
the universe reheats and the energy in gravitational waves redshifts
like radiation thereafter.

%%%%%%%%%%%%%%%%%%%%%%%%%
\begin{figure}[t]
\vspace{-4mm}
\begin{center}
\includegraphics[width=10cm,angle=-90]{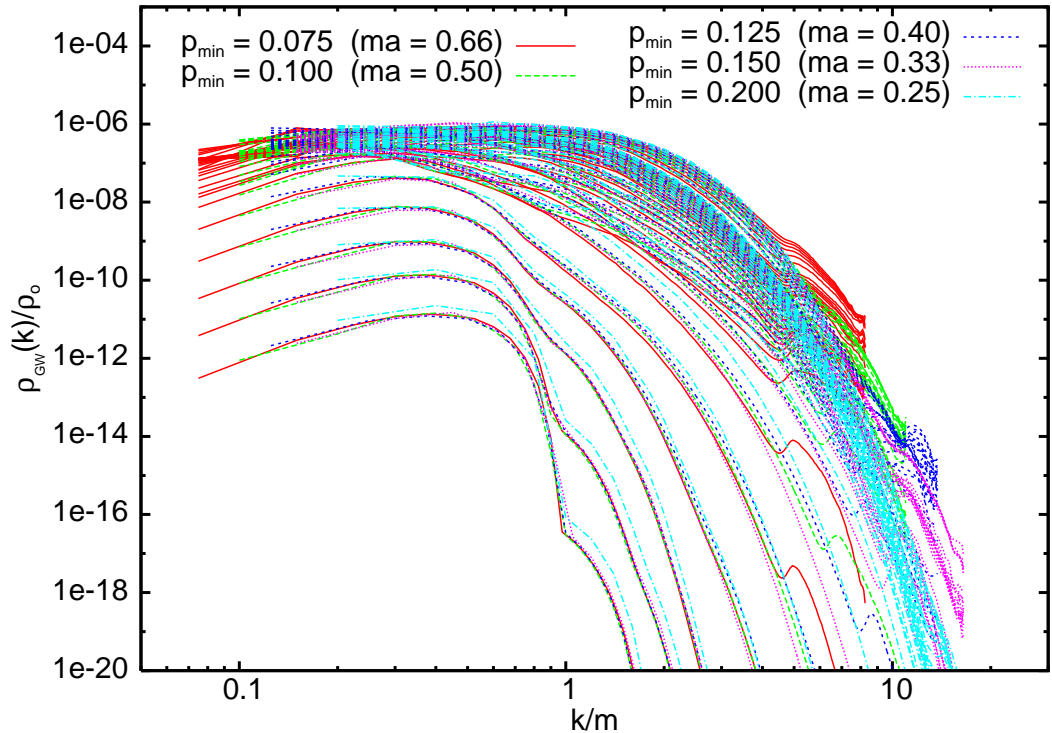}
\end{center}
\vspace*{-5mm}
\caption{We show here the comparison between the power spectrum of
gravitational waves obtained with increasing lattice resolution, to
prove the robustness of our method. The different realizations are
characterized by the %number of lattice points (N), 
the minimum 
lattice momentum (p$_{\rm min}$) and the lattice spacing (ma). The
growth is shown in steps of $m\Delta t = 1$ up to $mt = 30$, and 
% for the lower
then in and $m\Delta t = 5$ steps up to $mt = 60$.}
\label{fig4}
\vspace*{-3mm}
\end{figure}
%%%%%%%%%%%%%%%%%%%%%%%%%

To compute the power spectrum per logarithmic frequency
interval in GW, $\Omega_{_{\rm GW}}(f)$, we just have to use~(\ref{OmegaFraction}).
%\begin{equation}
%\indent \Omega_{_{\rm GW}}(k)={k^3\over2\pi^2}\,{\rho_{_{\rm GW}}(k)
%\over\rho_c}\,, 
%\end{equation}
%Moreover, since gravitational waves below Planck scale remain decoupled 
%from the plasma immediately after
%production, 
We can evaluate the power spectrum today from that
obtained at reheating by converting the wavenumber $k$ into
frequency $f$. Simply using Eq.~(\ref{redshift}), with 
$g_{{r},s}/g_{0,s}\sim 100$, $g_{{r},s}\,\sim\,g_{r}$ and 
$a_e \sim a_*$, then
\begin{equation}
 f=6\times10^{10}\,{\rm Hz}\,{k\over\sqrt{H\,M_p}}
=5\times10^{10}\,{\rm Hz}\,{k\over m}\,\lambda^{1/4}\,.
\end{equation}
We show in Fig.~\ref{fig4} the power spectrum of gravitational waves
as a function of (comoving) wavenumber $k/m$. We have used different
lattices in order to have lattice artifacts under control, specially
at late times and high wavenumbers. We made sure by the choice of
lattice size and spacing (i.e. $k_{\rm min}$ and $k_{\rm max}$) that
all relevant scales fitted within the simulation. Note, however, that
the lower bumps are lattice artifacts, due to the physical cutoff
imposed at the initial condition, that rapidly disappear with time. We
have also checked that the power spectrum of the scalar fields follows
turbulent scaling after $mt \sim {\cal O}(100)$, and we can thus
estimate the subsequent evolution of the energy density distributions
beyond our simulations.

\section{Lattice simulations}

The problem of determining the time evolution of a quantum field
theory is an outstandingly difficult problem. In some cases only a few
degrees of freedom are relevant or else perturbative techniques are
applicable. However, in our particular case, our interests are focused
on processes which are necessarily non-linear and non-perturbative and
involve many degrees of freedom. The presence of gravitational fields
just contributes with more degrees of freedom, but does not complicate 
matters significantly.

%The lattice formulation allows a first principles approach to
%non-perturbative quantum field theory. The existing powerful lattice
%field theory numerical methods rest on the path integral formulation
%in Euclidean space and the existence of a probability measure in field
%space~\cite{Wilson}. However, the problem we are interested in is a
%dynamical process far from equilibium, and the corresponding Minkowski
%path integral formulation is neither mathematically well founded nor
%appropriate for numerical studies. There are a series of alternative
%non-perturbative methods which different research groups have used to
%obtain physical results in situations similar to ours. These include
%Hartree's approximations~\cite{Cooper} to go beyond perturbation
%theory or large N techniques~\cite{Vega,Baacke}. It is clear that it
%is desirable to look at this and similar problems with all available
%tools. 
In the present paper we will use the so called classical approximation
to deal with the problem. It consists of
replacing the quantum evolution of the system by its classical
evolution, for which there are feasible numerical methods
available. The quantum nature of the problem remains in the stochastic
character of the initial conditions. This approximation has been used
with great success by several groups in the
past~\cite{classical,tachyonic}. The advantage of the method is that
it is fully non-linear and non-perturbative. %, allows the use of gauge
Our approach is to discretize the classical
equations of motion of all fields in both space and time. The
time-like lattice spacing $a_t$ must be smaller than the spatial one
$a_s$ for the stability of the discretized equations. In addition to
the ultraviolet cut-off one must introduce an infrared cut-off by
putting the system in a box with periodic boundary conditions. % We
%have studied $64^3$, $128^3$ and $256^3$ lattices.  Computer memory
%and CPU resources limit us from reaching much bigger lattices.
%Nonetheless, there are a number of
%checks one can perform to ensure that our results are physical and are
%not biased, within errors, by the approximations introduced, see 
%Fig.~\ref{fig4}. Our problem has several physical scales which control
%different time-regimes and observables. Thus, it is not always an easy
%matter to place these scales in the window defined by our ultraviolet
%and infrared cut-offs. %For example, in addition to the Higgs and
%inflaton mass there is a scale $M$ associated to the inflaton velocity
%which is particularly relevant in determining the bubble sizes and
%collisions. 
%Then, when we want to study a stage of the evolution in
%particular, we make the selection of the volume and cutoff most
%suitable. Since in this paper we are more interested in understanding the
%phenomenon of GW production, rather than concentrating in a particular
%model, our attitude has been to modify the parameters of the model in
%order to sit in a region where our results are insensitive to the
%cut-offs. This is no doubt a necessary first step to determine the
%requirements and viability of the study of any particular model. In
%particular, 
In this paper we have thouroughly studied a model with
$g^2=2\lambda=1/4$, but we have checked that other values of the
parameters do not change our results significantly.

\subsection{Initial conditions}

%Our approach to the dynamics of the system is to assume that the
%leading effects under study can be well-described by the classical
%evolution of the system. The justification of this point, as explained
%in the previous section, rests upon the fast quantum evolution of the
%long wavelength components of the Higgs field during the initial
%stages after the critical point. All the other degrees of freedom will
%evolve slowly from their initial quantum vacuum state. For the Higgs
%field, the leading behaviour is the exponential growth of those modes
%having negative effective mass-squared. The quantum evolution of such
%modes drives the system into a quasi-classical regime. It is essential
%that this regime is reached before the non-linearities couple all
%degrees of freedom to each other and questions like back-reaction
%start to affect the results. It is then assumed that it is the
%essentially classical dynamics of that field what matters, and that
%all the long-wavelength components of the inflaton and the gauge
%fields produced by the interaction with the Higgs field behave also as
%classical fields. Of course, this can hardly be the case for shorter
%wavelengths which stay in a quantum state with low occupation numbers.
%However, as we can see in Fig.~\ref{fig4}, for the range of times
%studied in this paper, the effect of shorter wavelengths is relatively
%small, and the spectrum of modes remains always dominated by
%long-wavelengths. 

The initial conditions of the fields follow the prescription from 
Ref.~\cite{symmbreak}. The Higgs modes $\phi_k$ are solutions of the 
coupled evolution equations, which can be rewritten as
$\phi''_k+(k^2-\tau)\phi_k=0$, with $\tau=M(t-t_c)$ and
$M=(2V)^{1/3}m$. The time-dependent Higgs mass follows from the
initial inflaton field homogeneous component, $\chi_0(t_i) =
\chi_c(1-Vm(t_i-t_c))$ and $\dot\chi_0(t_i)=-\chi_cVm$. The Higgs
modes with $k/M > \sqrt{\tau_i}$ are set to zero, while the rest are
determined by a Gaussian random field of zero mean distributed
according to the Rayleigh distribution
\begin{equation}
P(|\phi_k|)d|\phi_k|d\theta_k = \exp\left(-{|\phi_k|^2\over\sigma_k^2}
\right)\,{d|\phi_k|^2\over\sigma_k^2}\,{d\theta_k\over2\pi}\,,
\end{equation}
with a uniform random phase $\theta_k\in[0,2\pi]$ and dispersion given 
by $\sigma_k^2 \equiv |f_k|^2 = P(k,\tau_i)/k^3$, where $P(k,\tau_i)$ is the power 
spectrum of the initial Higgs quantum fluctuations, computed in the linear 
approximation in the background of the homogeneous inflaton.
%In the region of low momentum modes it is well approximated by
%\begin{equation}\label{Papp}
%\indent 
%2kP_{\rm app}(k,\tau_i) = k^3\left(1+A(\tau_i)\,k^2\,e^{-B(\tau_i)\,k^2}\right)\,,
%\end{equation}
%where $A(\tau_i)$ and $B(\tau_i)$ are parameters extracted from a fit
%of this form to the exact power spectrum given in paper~I.
In the
classical limit, the conjugate momentum $\dot\phi_k(\tau)$ 
is uniquely determined as $\dot\phi_k(\tau) = F(k,\tau)\phi_k(\tau)$, with
$F(k,\tau) = {\rm Im}(if_k(\tau)\dot f_k^*(\tau))/|f_k(\tau)|^2$, see
Ref.~\cite{symmbreak}. 

The rest of the fields (the inflaton non-zero modes and the
gravitational waves), are supposed to start from the vacuum, and
therefore they are semiclassically set to zero initially in the
simulations. Their coupling to the Higgs modes will drive their
evolution, giving rise to a rapid (exponential) growth of the GW and
inflaton modes. Their subsequent non-linear evolution will be well
described by the lattice simulations. In the next subsections we will 
describe the different evolution 
stages found in our simulations.

\subsection{Tachyonic growth}

In this subsection we will compare the analytical estimates with our
numerical simulations for the initial tachyonic growth of the Higgs
modes and the subsequent growth of gravitational waves. The first
check is that the Higgs modes grow according to Ref.~\cite{symmbreak}.
There we found that 
\begin{equation}\label{kphi2}
k|\phi_k(t)|^2 \simeq v^2\,A(\tau)\,e^{-B(\tau)k^2}\,, 
\end{equation}
with $A(\tau)$ and $B(\tau)$ are given, for $\tau>1$, as
%\begin{equation}\label{ABT}
%\indent 
$A(\tau) = {\pi^2(1/3)^{2/3}\over2\Gamma^2(1/3)}\,{\rm Bi}^2(\tau)\,,
$ and
$B(\tau) = 2(\sqrt\tau - 1)$,
%\end{equation}
where ${\rm Bi}(z)$ is the Airy 
function of the second kind. %Indeed, 
We have checked 
%see in Fig.~\ref{HigSpec} 
that the initial growth, from $mt=6$ to
$mt=10$, follows precisely the analytical expression.%, once taken into
%account that in Eq.~(\ref{kphi2}) the wavenumber $k$ and time $\tau$ are
%given in units of $M=(2V)^{1/3}m$.

The comparison between the tensor modes $h_{ij}(k,t)$ and the
numerical results is somewhat more complicated. We should first
compute the effective anisotropic tensor T$_{ij}({\bf
k},t)$~(\ref{GWfakeSource}) from the gradients of the Higgs field (those
of the inflaton are not relevant during the tachyonic growth), as
follows,
\begin{equation}
\tilde\Pi_{ij}({\bf k},t) = \int {d^3{\bf x}\,e^{-i{\bf k}{\bf x}}
\over(2\pi)^{3/2}}
\left[\nabla_i\phi^a\,\nabla_j\phi^a({\bf x},t) 
%-\frac{1}{3}(\nabla\phi)^2\delta_{ij}
\right]\,,
\end{equation}
where $\nabla_i\phi^a({\bf x},t) = \int {d^3{\bf q}\over(2\pi)^{3/2}}
\,iq_i\,\tilde{\phi^a}({\bf q},t)\,e^{-i{\bf q}{\bf x}}$. 
After performing the integral in ${\bf x}$ and using the delta
function to eliminate ${\bf q}'$, we make a change of variables
${\bf q}\to {\bf q}+{\bf k}/2$, and integrate over ${\bf q}$. 
Finally, with the use of $\tilde\Pi_{ij}({\bf k},t)$, we can compute 
the tensor fields,
\begin{eqnarray}
h_{ij}({\bf k},t) = (16\pi G)\,\int_0^t\,dt'\,{\sin k(t-t')\over k}\,
\tilde\Pi_{ij},\hspace{.6cm}
\partial_0 h_{ij}({\bf k},t) = (16\pi G)\,\int_0^t\,dt'\,\cos k(t-t')\,
\tilde\Pi_{ij}.
\end{eqnarray}
Using the analytic solutions one can perform the
integrals and obtain expressions that agree surprisingly well with
the numerical estimates. This allows one to compute the density in
gravitational waves, $\rho_{_{\rm GW}}$, at least during the initial 
tachyonic stage in terms of analytical functions, and we reproduce
the numerical results.
We will now compare these with the analytical estimates.
The tachyonic growth is dominated by the faster-than-exponential
growth of the Higgs modes towards the true vacuum.
The (traceless) anisotropic strees tensor $\Pi_{ij}$ grows rapidly to
a value of order $k^2|\phi|^2 \sim 10^{-3}\,m^2v^2$, which gives a
tensor perturbation 
\begin{equation}
\left|h_{ij} h^{ij}\right|^{1/2} \sim 16\pi
G v^2 (m\Delta t)^2 10^{-3}\,,
\end{equation}
and an energy density in GW, 
\begin{equation}
\rho_{_{\rm GW}}/\rho_0 \sim 64\pi G v^2\,(m\Delta t)^2 10^{-6} \sim Gv^2\,, 
\end{equation}
for $m\Delta t \sim 16$.  In the case at hand, with $v=10^{-3}\,M_P$, we
find $\rho_{_{\rm GW}}/\rho_0 \sim 10^{-6}$ at symmetry breaking, which
coincides with the numerical simulations at that time, see
Fig.~\ref{fig3}. 

As shown in Ref.~\cite{symmbreak}, the spinodal instabilities grow
following the statistics of a Gaussian random field, and therefore one
can use the formalism of~\cite{BBKS} to estimate the number of peaks
or lumps in the Higgs spatial distribution just before symmetry
breaking.  As we will discuss in the next section, these lumps will
give rise via non-linear growth to lump invagination and the formation
of bubble-like structures with large density gradients, expanding at
the relativistic speeds and colliding among themselves giving rise to a
large GWB. The size of the bubbles upon collision is essentially
determined by the distance between peaks at the time of symmetry
breaking, but this can be computed directly from the analysis of
Gaussian random fields, as performed in Ref.~\cite{symmbreak}.
This analysis works only for the initial (linear) stage before symmetry
breaking.  Nevertheless, we expect the results to extrapolate to later
times since once a bubble is formed around a peak, it remains there at
a fixed distance from other bubbles. This will give us an idea of the
size of the bubbles at the time of collision.

\subsection{Bubble collisions}

The production of gravitational waves in the next stage proceeds
through `bubble' collisions. In Ref.~\cite{tachyonic} we showed
%explicitly that symmetry breaking is not at all a homogeneous
%process. 
that during the symetry breaking, the Higgs field develops
lumps whose peaks grow up to a maximum value $|\phi|_{\rm max}/v =
4/3$, and then decrease creating approximately
spherically symmetric bubbles, with ridges that remain above $|\phi| =
v$. Finally, neighboring bubbles collide and high momentum modes are
induced via field inhomogeneities. Since 
initially only the Higgs field sources the anisotropic stress-tensor
$\Pi_{ij}$, then we expect the formation of structures 
%(see section IV.A) 
in the spatial distribution of the GW energy density 
%the tensor metric perturbation, 
correlated with the Higgs lumps. 
%In section IV, %we will show the explicit form
%of the structures developed in the spatial distribution of
%$\rho_{_{\rm GW}}$ related with the first collisions among the
%bubble-like structures of the Higgs field. 
%we will show simultaneously the evolution of both the Higgs'
%spatial distribution when the first bubbles start colliding, 
%and of the corresponding structures in the GW energy density 
%$\rho_{_{\rm GW}}$, expliciting the correlation between them.
%We leave for a
%forthcoming publication the details of an analytical formalism
%describing the formation and subsequent evolution of such GW
%structures.
In this sub-section we will give an estimate of the
burst in GW produced by the first collisions of the Higgs bubble-like
structures.

%The dependence of the $h_{ij}$ tensor on the gradient of the
%Higgs field, see~Eq.(\ref{GWeq}), is responsible of the formation of
%those structures in the energy density spatial distribution of the
%GWB.

As for the collision of vaccum bubbles in first
order phase transitions~\cite{KosowskyTurner}, we can give a simple 
estimate of the order of magnitude of the energy fraction radiated 
in the form of gravitational waves when two Higgs bubble-like structures 
collide. A similar stimation is indeed presented 
in~\cite{Felder:2006cc,DufauxGW}. In general, the problem of two 
colliding bubbles has several 
time and length scales: the duration
of the collision, $\Delta t$; the bubbles' radius $R$ at the moment of
the collision; and the relative speed of the bubble walls. %In section
%IV.B we found that 
The typical size of bubbles upon collisions, is of the order of $R\approx 10m^{-1}$, 
while the growth of the bubble's wall is relativistic, see Ref.~\cite{symmbreak}. 
Then we can assume than the time scale
associated with bubble collisions is also $\Delta t\,\sim\,R$.
Assuming the bubble walls contain most of the energy density, 
it is expected that the asymmetric collisions will copiously produce GW.

Far from a source that produces gravitational radiation, the dominat
contribution to the amplitude of GW is given by the acceleration of
the quadrupole moment of the Higgs field distribution. Given
the energy density of the Higgs field, $\rho_{\rm H}$, we can compute
the (reduced) quadrupole moment of the Higgs field spatial
distribution, $Q_{ij} = \int d^3x\,(x_i x_j -
x^2\delta_{ij}/3)\,\rho_{\rm H}(x)$, such that the amplitude of the
gravitational radiation, in the TT gauge, is given by 
$h_{ij}\,\sim\,(2G/r)\ddot Q_{ij}$. A significant amount of energy can be
emitted in the form of gravitational radiation whenever the
quadrupole moment changes significantly fast: through the bubble
collisions in this case. The power carried by these waves can be
obtained via~(\ref{rhoTotal}) as
\begin{eqnarray}
 P_{_{\rm GW}} = \frac{G}{8\pi}\int d\Omega\,\left\langle  
\dddot{Q}_{ij}\dddot{Q}^{ij}\right\rangle\,. 
\end{eqnarray} 
Omitting indices for simplicity, as the power emitted in gravitational
waves in the quadrupole approximation is of order $P_{_{\rm
GW}}\,\sim\,G(\dddot{Q})^2$, while the quadrupole moment is of order
$Q \sim R^5\rho_{\rm H}$, we can estimate the power emitted in GW
upon the collision of two Higgs bubbles as
\begin{equation}
 P_{_{\rm GW}}\,\sim\,G\left(\frac{R^5\rho}{R^3}\right)^2\,
\sim\,G\rho_{\rm H}^2\,R^4
\end{equation} 
The fraction of energy density carried by these waves, $\rho_{_{\rm
GW}}\,\sim\,P_{_{\rm GW}}\Delta t/R^3\,\sim\,P_{_{\rm
GW}}/R^2\,\sim\,G\rho_{\rm H}^2\,R^2$, compared to that of the initial
energy stored in the two bubble-like structures of the Higgs field,
will be $\rho_{_{\rm GW}}/{\rho_{\rm H}} = G\rho_{\rm H}R^2$.  Since
the expansion of the universe is negligible during the bubble
collision stage, the energy that drives inflation, $\rho_0\,\sim\,
m^2v^2$, is transferred essentially to the Higgs modes during
preheating, within an order of magnitude, see Fig.~\ref{fig3}. Thus,
recalling that $R\,\sim 10m^{-1}$, the total fraction of energy in GW
produced during the bubble collisions to that stored in the Higgs
lumps formed at symmetry breaking, is given by
\begin{eqnarray}\label{GWbubbles}
 \frac{\rho_{_{\rm GW}}}{\rho_0}\,\sim 0.1\,G\rho_0\,R^2\,
\sim\,(v/M_p)^2\,,
\end{eqnarray} 
giving an amplitude which is of the same order as is observed in the
numerical simulations, see Fig.~\ref{fig3}. Of course, an exhaustive
analytical treatment of the production of GW during this stage of
bubble collisions remains to be done, but we leave it for a future
publication.

%%%%%%%%%%%%%%%%%%%%%%%%%
\begin{figure}[t]%[t]
%\vspace{5mm}
\begin{center}
\includegraphics[width=10cm,angle=-90]{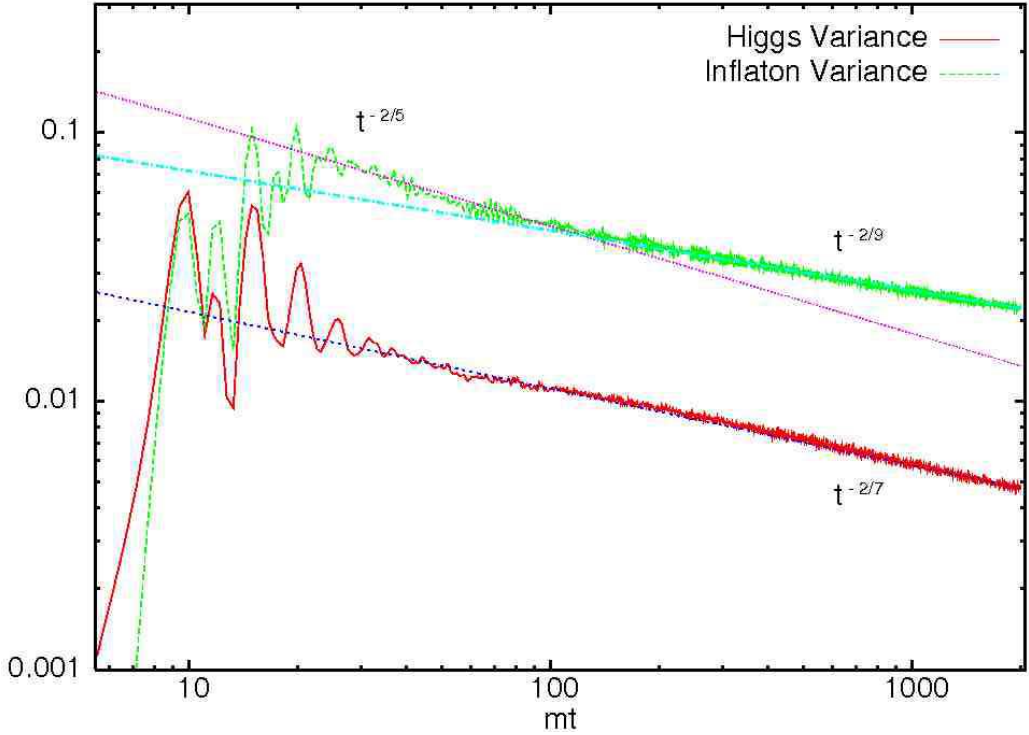}
\end{center}
\vspace*{-5mm}
\caption{Variance of the Inflaton and the Higgs field as a function of
time, the former normalized to its critical value, the latter
normalized to its \textit{v.e.v.}. As expected in a turbulent regime,
these variances follow a power law $\sim t^{-2p}$ with $p$ a certain
critical exponent, although the slope of the Inflaton's variances
evolves in time. The curves are produced from an average over 10 
different statistical realizations.}
\label{variances_fig}
\vspace*{-3mm}
\end{figure}
%%%%%%%%%%%%%%%%%%%%%%%%%

\subsection{Turbulence}

The development of a turbulent stage is expected from the point of
view of classical fields, as turbulence usually appears whenever there
exists an active (stationary) source of energy localized at some scale
$k_{\rm in}$ in Fourier space. 
%As first pointed out by
%Ref.~\cite{MichaTkachev}, in reheating scenarios the coherently
The oscillating inflaton zero-mode plays the role of the pumping-energy
source, acting at a well defined scale $k_{\rm in}$ in Fourier space,
given by the frequency of the inflaton oscillations. %Thus, the
%inflaton zero-mode pumps energy into the rest of the fields that
%couple to it as well as into the non-zero modes of the inflaton field
%itself. 
Apart from $k_{\rm in}$, there is no other scale in Fourier
space where energy is accummulated, dissipated and/or infused. So, as
turbulence is characterized by the transport of some conserved
quantity, energy in our case, we should expect a flow of energy from
$k_{\rm in}$ towards higher (direct cascade) or smaller (inverse
cascade) momentum modes. In typical turbulent regimes of classical
fluids, there exits a sink in Fourier space, corresponding to that
scale at which the (direct) cascade stops and energy gets
dissipated. However, in our problem there is no such sink so that the
transported energy cannot be dissipated, but instead it is used to
populate high-momentum modes. For the problem at hand, there exists a
natural initial cut-off $k_{\rm out} \sim \lambda^{1/2}v,$ such that
only long wave modes within $k<k_{\rm out}$, develop the spinodal
instability. Eventually, after the tachyonic growth has ended and the
first Higgs' bubble-like structures have collided, the turbulent
regime is established. Then the energy flows from small to greater
scales in Fourier space, which translates into the increase of $k_{\rm
out}$ in time.

%%%%%%%%%%%%%%%%%%%%%%%%%
\begin{figure}[t]%[t]
%\vspace{5mm}
\begin{center}
\includegraphics[width=7cm,height=11.5cm,angle=-90]{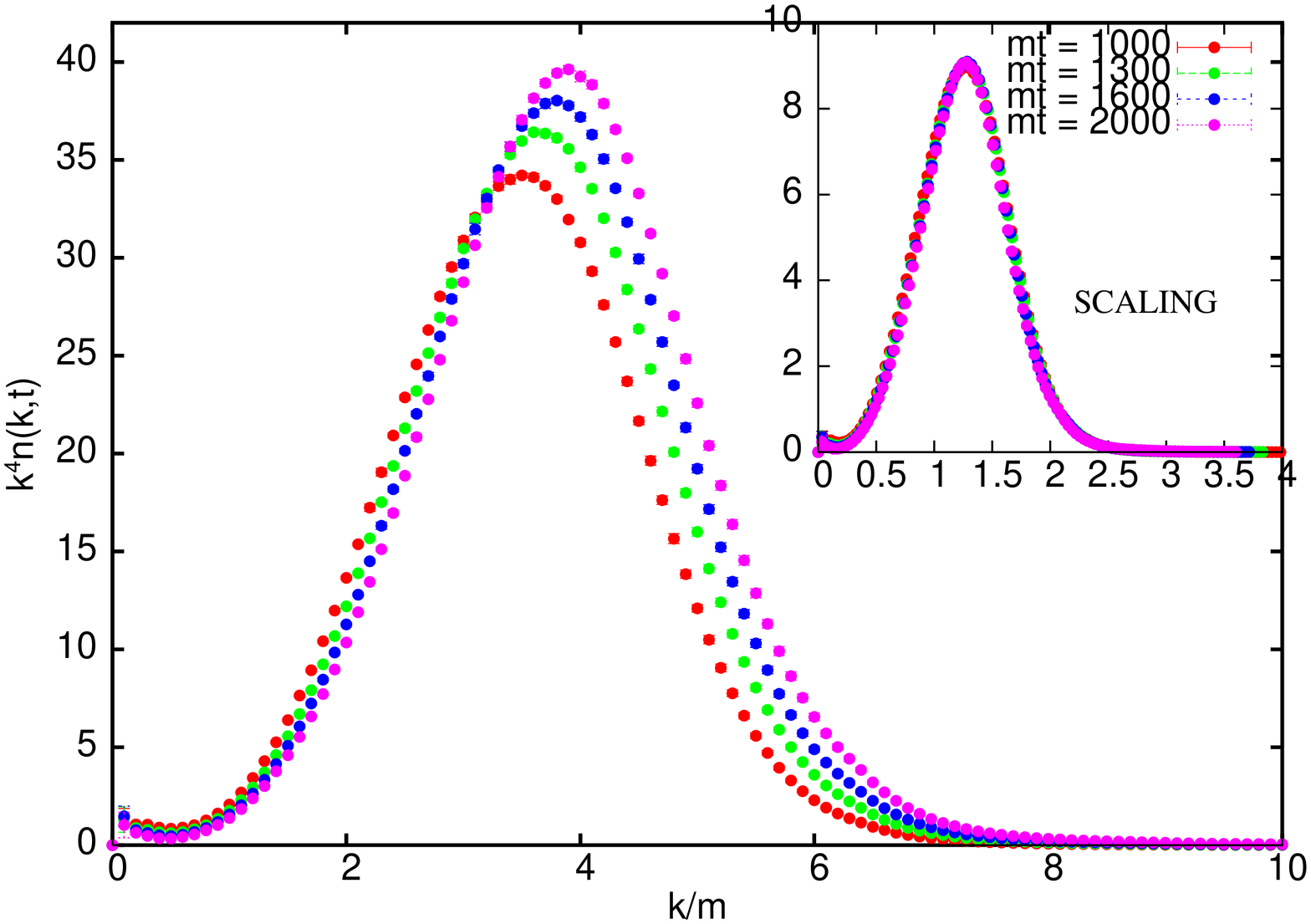}
\includegraphics[width=7cm,height=11.5cm,angle=-90]{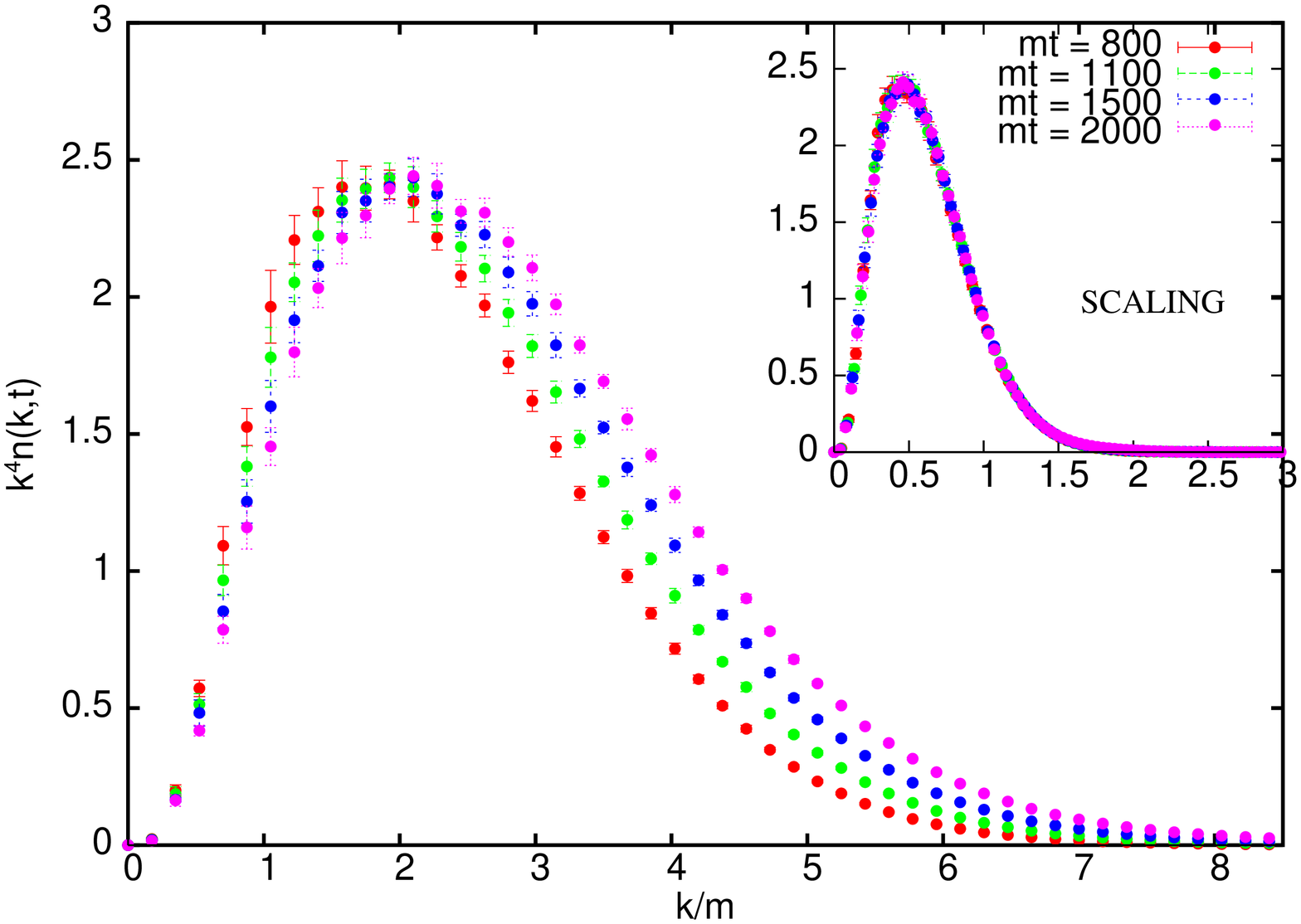}
\end{center}
\vspace*{-5mm}
\caption{Some snapshots of the evolution of the spectral particle
occupation numbers of the Higgs and the Inflaton fields at different 
times, each averaged over 10 statistical realizations. We multiply them by $k^4$
so we can see better the scaling behaviour. In the upper right corner,
we plot the inverse relation of~(\ref{selfSimilar}), $n_0(k t^{-p})
= t^{\gamma p}n(k,t)$, also averaged over 10 realizations for each
time. The scaling behaviour predicted by wave kinetic turbulent
theory~\cite{MichaTkachev}, is clearly verified.}
\label{higgsScaling_fig}
\vspace*{-3mm}
\end{figure}
%%%%%%%%%%%%%%%%%%%%%%%%%

%In Ref.~\cite{magnetic} we already accounted for the turbulent stage
%reached in a hybrid model with gauge fields. However, we don't
%consider gauge fields here, so the number of degrees of freedom is
%different from that of Ref.~\cite{magnetic} and, therefore, the turbulent
%dynamics of the Inflaton and the Higgs fields should be different. In
%particular, 
When the turbulence has been fully established, if the
wave (kinetic) turbulence regime of the fields' dynamics is valid, the
time evolution of the variance of a turbulent field $f({\bf x},t)$,
should follow a power-law-like scaling~\cite{MichaTkachev}
\begin{equation}
\label{variances}
 \mathrm{Var}(f(t)) = \left\langle f(t)^2\right\rangle -
\left\langle f(t)\right\rangle^2 \propto \ t^{-2p} \,,
\end{equation}
with $p = 1/(2N-1)$ and $N$ the number of scattering fields in a
`point-like collision'. %In fact, such time behaviour corresponds only
%to the case of the so called \textit{free turbulence}, when the energy
%stored in the pumping source is subdominant to the energy in the
%turbulent fields. In our case, this condition is reached very soon
%after the symmetry breaking, so we don't expect a significant stage of
%\textit{driven turbulence}, which would make the variance to increase
%(Only the inflaton seems to increase its variance between $mt=10$ and
%$mt=30$, but it is not very pronounced). 
In Fig.~\ref{variances_fig} we have plotted the time evolution of the 
variances of the Inflaton
$\chi$ and of the Higgs modulus $\phi = \sqrt{\sum_a\phi_a^2}$, and
fitted the data with a power-law like~(\ref{variances}), obtaining
\begin{flushleft}
\begin{tabular}{rrll}
\vspace{.2cm} \,\,\,\, & Inflaton: & $p^{-1}_{_I} = 5.1 \pm 0.2$, &
\,\,\,[35:85] \\ \vspace{.2cm} \,\,\,\,& Inflaton: & $p^{-1}_{_I} =
9.03 \pm 0.03$, & \,\,\,[350:2000] \\ \vspace{.1cm} \,\,\,\, & Higgs:
& $p^{-1}_{_H} = 7.02 \pm 0.01$, & \,\,\,[50:2000]
\end{tabular}
\end{flushleft}
where the last brackets on the right correspond to the range in time 
(in units of $m^{-1}$) for which we fitted the data. As can be seen in
Fig.~\ref{variances_fig}, the slope of the Higgs field (in
logarithmic scale), $2p_{_H}\,\sim\,2/7$, remains approximately
constant in time, corresponding to a 4-field dominant interaction.
However, the slope of the Inflaton's variance increases in time, 
i.e. the critical exponent $p_{_I}$ of the Inflaton decreases, 
until it reaches a stationary stage at $mt\sim100$. 
%Since $p_{_I}$ is related to the number $N$ of fields interacting in a
%collision, if there was a change from one dominant multi-field
%interaction to another, this should produce a time-dependent effective
%$p_{_I}$, as seen in Fig.~\ref{variances_fig}. However, 
We will not try to explain here the origin of such an effective 
critical exponents
as extracted from the simulations. We will just stress that we 
have checked the robustness of those values under different lattice 
configurations ($N,p_{\rm min}$) and different statistical realizations.
%, discarding this way a possible lattice artefact effect. 
%As we will see, the critical exponents $p$ determines the speed with 
%which the turbulent particle distribution moves over momentum space, 
%so this is a crucial parameter. %Moreover, although both the classical 
%modes of the Inflaton
%and the Higgs contribute to the production of GW, the Inflaton's
%occupation numbers decrease faster than those of the Higgs so, after a
%given time, only the Higgs' modes remain as the main source of GW.
Actually, when turbulence has fully developed, it is expected that the
distribution function of the classical turbulent fields, the inflaton
and the Higgs here, follow a self-similar
evolution~\cite{MichaTkachev}
\begin{eqnarray}
\label{selfSimilar}
 n(k,t) = t^{-\gamma\,p}n_0(k\,t^{-p})\,,
\end{eqnarray}
with $p$ the critical exponent of the fields' variances and $\gamma$ a
certain factor $\sim O(1)$, which depends on the type of turbulence
developed. Looking at~(\ref{selfSimilar}), we see that the exponent 
$p$ determines the speed of the particles' distribution in momentum 
space: given a
specific scale $k_c$ %such that, for example, the occupation number has
%a maximum, 
that scale evolves in time as $k_c(t) = k_c(t_0)(t/t_0)^{p}$. 
In the simulations, we have seen that the evolution of the Higgs 
occupation number follows
Eq.~(\ref{selfSimilar}) with $p \approx 1/7$, as expected from the
Higgs variance, and $\gamma \approx 2.7$. Whereas the evolution of the
Inflaton occupation number follows~(\ref{selfSimilar}) even more
accurately than the Higgs, with an ``effective" exponent 
(once the asymptotic regime is achieved) $p \approx
1/5$, and $\gamma \approx 3.9$. %Since the slope of the inflaton's
%variance changes in time, the value of the exponents of the inflaton's 
%scaling relation will require further investigation. However, despite 
%this time evolution of the Inflaton variance, Eq.~(\ref{selfSimilar})
%is very well fulfilled by the Inflaton with the given effective 
%exponents. So we can perfectly obtain the universal $n_0(k)$ function 
%for the Inflaton as well as for the Higgs.
%%%%%%%%%%%%%%%%%%%%%%%%%
\begin{figure}[t]%[t]
%\vspace{5mm}
\begin{center}
\includegraphics[width=10cm,angle=-90]{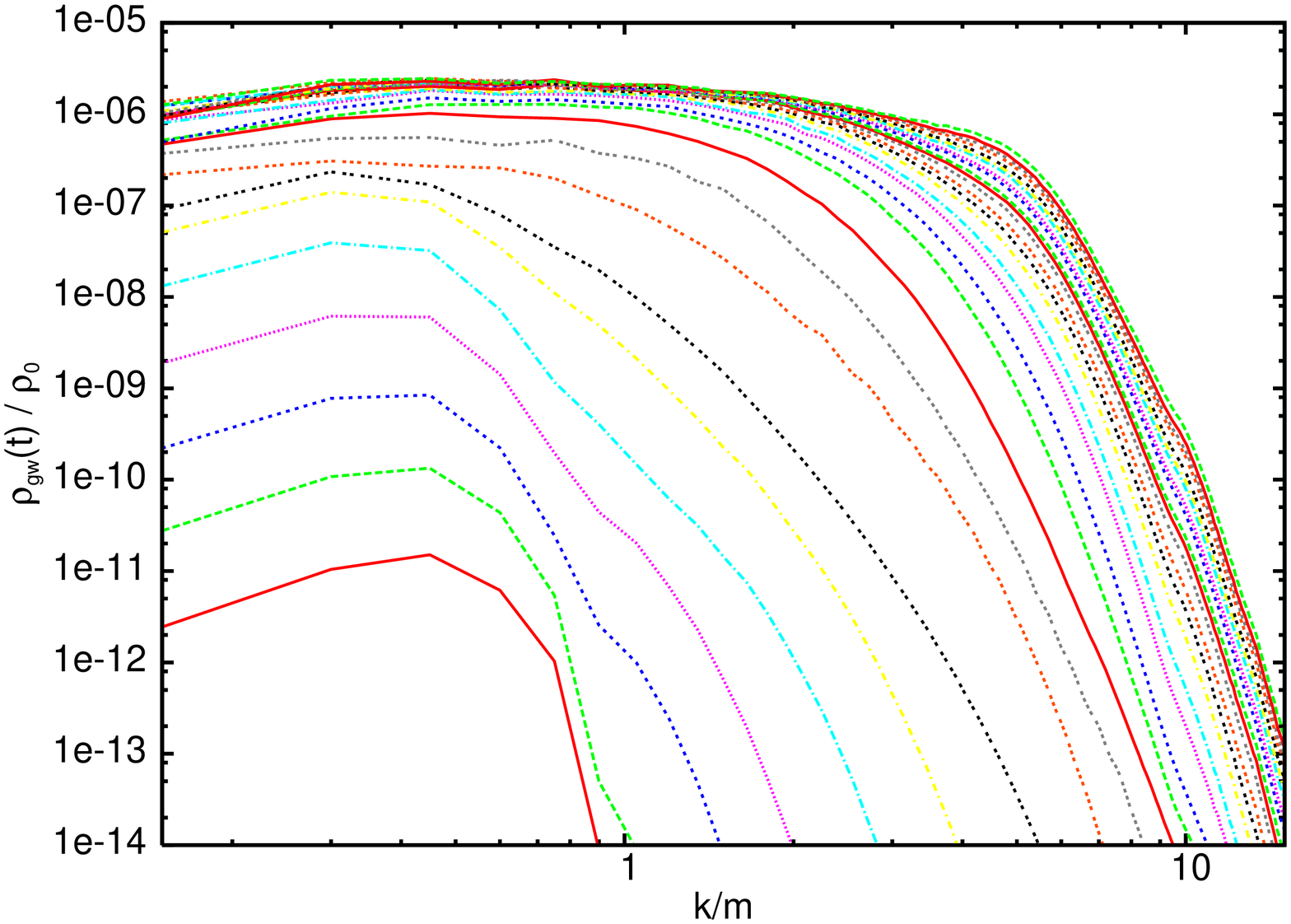}
\end{center}
\vspace*{-5mm}
\caption{Time evolution of the GW spectra from mt = 6 to mt = 2000. 
The amplitude of the spectra seems to saturate after mt $\sim 100$, 
although the high momentum tail still moves slowly to higher values 
of $k$ during the turbulent stage.}
\label{fig10}
\vspace*{-3mm}
\end{figure}
%%%%%%%%%%%%%%%%%%%%%%%%%
In Figs.~\ref{higgsScaling_fig} % and~\ref{inflatonScaling_fig}
we have plotted the occupation numbers of the Higgs and the Inflaton, also
inverting the relation of Eq.~(\ref{selfSimilar}) in order to extract
the \textit{universal} time-independent $n_0(k)$ functions of each
field. As shown in those figures, the distributions follow nicely the
expected scaling behaviour. %However, for the range of interest of $k$,
%there are small discrepancies of order 0.1-4\% (depending on $k$)
%among the different universal functions $n_0^{(i)}(k)$, as obtained 
%inverting Eq.~(\ref{selfSimilar}) at different times $mt_i$. 
The universal functions $n_0(k)$ plotted in Figs.~\ref{higgsScaling_fig} 
%and~\ref{inflatonScaling_fig} 
have been obtained from averaging over
ten statistical realizations for each time.

The advantage of the development of a turbulence behaviour is obvious:
it allows us to extrapolate the time evolution of the fields'
distributions till later times beyond the one we can reach with the
simulations. Moreover, the fact that the turbulence develops so early
after the tachyonic instability, also allow us to check for a long
time of the simulation, the goodness of the description of the
dynamics of the fields, given by the turbulent kinetic theory
developed in Ref.~\cite{MichaTkachev}. We have fitted the averaged
universal functions $n_0(k)$ with expressions of the form $k^4\,n_0(k)
= P(k)e^{-Q(k)}$, with $P(k)$ and $Q(k)$ polynomials in $k$. %, giving: 
%\begin{eqnarray}
%\begin{array}{lll}
%\label{Inflaton_universal}
%{\rm Inflaton:} & P(k) = 486.2k^3 & Q(k) = 6.39k \\
%\,&\,& \\
%\label{Higgs_universal}
%{\rm Higgs:} & P(k) = 2.96k^3 & Q(k) = 2.26k^2 - 3.18k
%\end{array}
%\end{eqnarray} 
There is no fundamental meaning associated with such a fit,
but it is very useful to have an analytical control over 
$n_0(k)$, since this allows us to track the time-evolution of 
$n(k,t)$ through
Eq.~(\ref{selfSimilar}). Actually, the classical regime of the
evolution of some bosonic fields ends when the system can be relaxed
to the Bose-Einstein distribution. Since we cannot reach that moment, 
we can at least estimate the
moment in which the initial energy density gets fully transferred to
the Higgs classical modes. Using Eq.(\ref{selfSimilar}) and the
fit %~(\ref{Higgs_universal})
to the universal $n_0(k)$ of the Higgs, 
we find that the initial energy density 
% for $n(k)\,\sim \,1$ 
is totally transfered to the Higgs when (in units $m=1$)
\begin{equation}
\rho_0 = {1\over4\lambda} =
\int {dk\over k}\,{k^3\over2\pi^2}\,k\,n(k,t) =
{7.565\over2\pi^2}\, t^{(4-\gamma)p}\,,
\end{equation}
where we have assumed that the Higgs' modes have energy $E_k(k,t) =
k\,n(k,t)$. In our case, with $\lambda=1/8$, the conversion of the
initial energy density into Higgs particles and therefore into
radiation is complete by $t \sim 6\times10^4m^{-1}$. Therefore, if we 
consider this value as a lower bound for the time that classical 
turbulence requires to end, we see that turbulence last for a very 
long time compared to the time-scale of the initial tachyonic and bubbly 
stages. Thus, if GW were significatively sourced during turbulence, 
one should take into account corrections from the expansion of the universe.

In Fig.~\ref{fig10}, we show the evolution of the GW spectra up to
times mt $= 2000$, for a lattice of (N,\,$p_{\rm min}$) = (128,\,0.15). It
is clear from that figure that the amplitude of the GW saturates to a
value of order $\rho_{_{\rm GW}}/\rho_0 \approx 2\cdot10^{-6}$. At mt
$\approx 50$, the maximum amplitude of the spectra has already reached
$\rho_{_{\rm GW}}/\rho_0 \approx 10^{-6}$, while at time mt $\approx 100$,
the maximum has only grown a factor of 2 with respect to mt $\approx
50$. From times mt $\approx 150$ till the maximum time we reached in
the simulations, mt = 2000, the maximum of the amplitude of the
spectrum does not seem to change significantly, slowly increasing from
$\approx 2\cdot10^{-6}$ to $\approx 2.5\cdot10^{-6}$. Despite this
saturation, we see in the simulations that the long momentum tail
of the spectrum keeps moving towards greater values. This displacement
is precisely what one would expect from turbulence, although it is
clear that the amplitude of the new high momentum modes never exceed
that of lower momentum. In order to disscard that this displacement
towards the UV is not a numerical artefact, one should further
investigate the role played by the turbulent scalar fields as a source
of GW. Here, we just want to remark that the turbulent motions of the
scalar fields, seem not to increase significatively anymore the total
amplitude of the GW spectrum. Indeed, in a recent paper
\cite{DufauxGW} where GW production at reheating is also considered,
it is stated that GW production from turbulent motion of classical
scalar fields, should be very supressed. That is apparently what we
observe in our simulations although, as pointed above, this issue
should be investigated in a more detailed way. Anyway, here we can
conclude that the expansion of the Universe during reheating in these
hybrid models, does not play an important role during the time of GW
production, and therefore we can be safely ignore it.

\section{Gravitational Waves from Chaotic Inflation}

The production of a relic GWB at reheating was first 
addressed by Khlebnikov and Tkachev (KT) in 
Ref.~\cite{TkachevGW}, both for the quadratic and quartic chaotic
inflation scenarios. %In these models, the long-wavelength part of the
%spectrum is dominated by the gravitational bremsstrahlung associated
%with the scattering of the extra scalar particles off the inflaton
%condensate, `evaporating' this way the inflaton particles. Using
%this fact, 
%KT estimated analytically the amplitude of the power
%spectra of GW for the low frequency end of the spectrum,
%corresponding to wavelengths of order the size of the horizon at
%rescattering. Moreover, KT also studied the GW power spectra
%numerically, although just for the massless inflaton case. 
Recently, chaotic scenarios were revisited in 
Ref.~\cite{EastherLim,EastherLim2}. Also very recently, 
Ref.~\cite{DufauxGW} studied in a very detail way, 
the evolution of GW produced at preheating in the case of a 
massless inflaton with an extra scalar field.

In Refs.~\cite{TkachevGW} and~\cite{EastherLim}, the procedure
to compute the GW from reheating relied on Weinberg's 
formula %for the energy carried by a weak gravitational radiative field
for flat space-time~\cite{Weinberg}. However, in chaotic models, the
expansion of the universe cannot be neglected during reheating, 
so Weinberg's formula can only be used in an approximated way, 
if the evolution of the universe
is considered as an adiabatic sequence of stationary
universes. %Rescaling fields by a conformal transformation, their
%evolution equations can be solved with a numerical integrator, while
%the evolution of the scale factor can be calculated
%analytically. Discretizing the time, the physical variables can be
%recovered from the conformal ones in each time step, thus allowing to
%compute the energy of gravitational waves in terms of the physical
%fields. 
In Ref.~\cite{GBFS}, however, we adopted a different approach 
that takes into account the expansion of the universe in a
self-consistent manner, and allows us to calculate at any time 
the energy density and power spectra of the GW produced 
at reheating (see section 2). 
%As explained in section III and applied to the case of 
%hybrid inflation in sections IV and V,
%we just solve numerically Eq.~(\ref{GWfakeEq}), 
%together with those eqs. of the other Bose
%fields and the scale factor, Eqs.~(\ref{inflatonEq})%,(\ref{scalarEq}) 
%and Eqs.(\ref{hubbleDotEq}),(\ref{hubbleEq}).
%Then, using the projector~(\ref{projector}) into the (Fourier transformed) 
%solution of Eq.(\ref{GWfakeEq}), we recover the TT d.o.f corresponding 
%to GW. This way, we can monitor the total energy
%density in GW using Eq.~(\ref{rhoTotal}), or track the evolution 
%of the power
%spectrum. 
Using our technique, we will show in this
section that we reproduce, for specific chaotic models, 
similar results to those of other authors. In particular, 
we adapted the publicly available LATTICEEASY 
code~\cite{LatticeEasy}, taking advantage of the
structure of the code itself, incorparating the evolution of 
Eq.~(\ref{GWeq}), together with the equations of the scalar fields, 
Eqs.~(\ref{inflatonEq}), % and (\ref{scalarEq}),
into the staggered leapfrog integrator routine.

% This way, we can solve
%at the same time the dynamics of the scalar and tensor fields, within
%the framework of an expanding FRW universe~Eqs.(\ref{hubbleDotEq}) and
%(\ref{hubbleEq}). 

Here we will concentrate only in an scenario with a massless 
inflaton $\chi$, either accompanied or not by an extra scalar field $\phi$.
%In the following, we 
%will describe the numerical results for GW production at reheating in s
Such scenarios are described by the potential
\begin{equation}
V(\chi,\phi) = {\lambda\over 4}\chi^4 + {1\over 2}g^2\chi^2\phi^2
\end{equation}
Rescaling the time by and the physical fields by a conformal transformation as
\begin{eqnarray}
\chi_c(\tau) = {a(\tau)\over a(0)}{\chi(\tau)\over\chi(0)},\hspace{.5cm}
\phi_c(\tau) = {a(\tau)\over a(0)}{\phi(\tau)\over\chi(0)},\hspace{.5cm}
d\tau = {a(\tau)\over a(0)}\chi(0)\sqrt{\lambda}\,dt\,,
\end{eqnarray}
then the equations of motion of the inflaton and of the extra scalar field,
Eq.~(\ref{inflatonEq}),
% and (\ref{scalarEq}), 
can be rewritten in terms of the conformal variables as
\begin{eqnarray}
\label{inflatonEqChao4}
&&\chi_c'' - \nabla^2\chi_c -\frac{a''}{a}\chi_c + 
(\chi_c^2 + q\phi_c^2)\chi_c  = 0 \\
\label{scalarEqChao4}
&&\phi_c'' - \nabla^2\phi_c -\frac{a''}{a}\chi_c + 
q\chi_c^2\phi_c = 0\,,
\end{eqnarray}
where the prime denotes derivative with respect to conformal time. Since
the universe expands as radiation-like in these scenarios,
$a(\tau) \sim \tau$, so the terms proportional to $a''/a$
in Eqs.~(\ref{inflatonEqChao4}) and (\ref{scalarEqChao4}) are soon 
negligible, as
explicitly checked in the simulations. Thanks to this, the
model is conformal to Minkowski.

The parameter $q \equiv g^2/\lambda$ 
controls the strength and width of the resonance.
For the case of a massless inflaton without an extra scalar field,
we just set $q = 0$ in Eq.~(\ref{inflatonEqChao4}) and
ignore Eq.~(\ref{scalarEqChao4}). However, in that case, fluctuations
of the inflaton also grow via parametric resonance. Actually, they
grow as if they were fluctuations of a scalar field coupled to the
zero-mode of the inflaton with effective couplig $q = g^2/\lambda =
3$, see Ref.~\cite{Greene}.
Following Refs.~\cite{TkachevGW} and~\cite{EastherLim}, we set
$\lambda = 10^{-14}$ and $q$ = 120. Since this case is also 
computed in~\cite{DufauxGW}, we can also compare our results 
with theirs. Moreover, we also present results for the pure 
$\lambda\chi^4$ model with no extra scalar field, a case 
only shown in Ref.~\cite{TkachevGW}.

%%%%%%%%%%%%%%%%%%%%%%%%%
\begin{figure}[t]%[b]
%\vspace{5mm}
\begin{center}
\includegraphics[width=9cm,angle=-90]{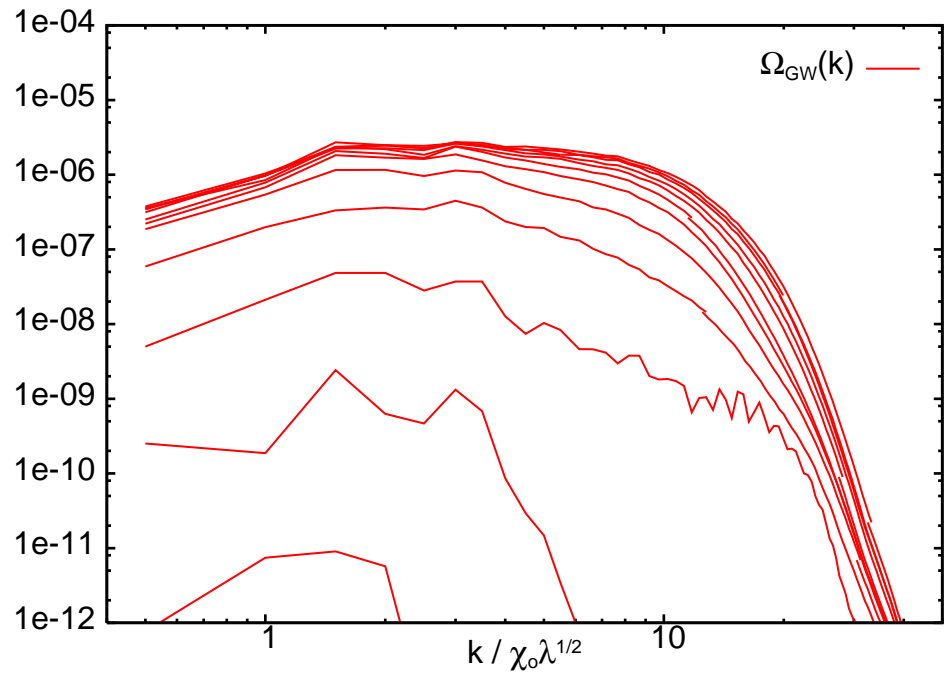}
\end{center}
\vspace*{-5mm}
\caption{The spectrum of the gravitational waves' energy density, 
for coupled case with $\lambda = 10^{-14}$ and $g^2/\lambda = 120$. 
The spectrum is shown accumulated up to different times during GW 
production, so one can see its evolution. At each time, it is normalized 
to the total instant density. This plot corresponds to a N = 128
lattice simulation, from $\tau = 0$ to $\tau = 240$.}
\label{fig6}
\vspace*{-3mm}
\end{figure}
%%%%%%%%%%%%%%%%%%%%%%%%%

\begin{figure}[t]%[t]
%\vspace{5mm}
\begin{center}
\includegraphics[width=9cm,angle=-90]{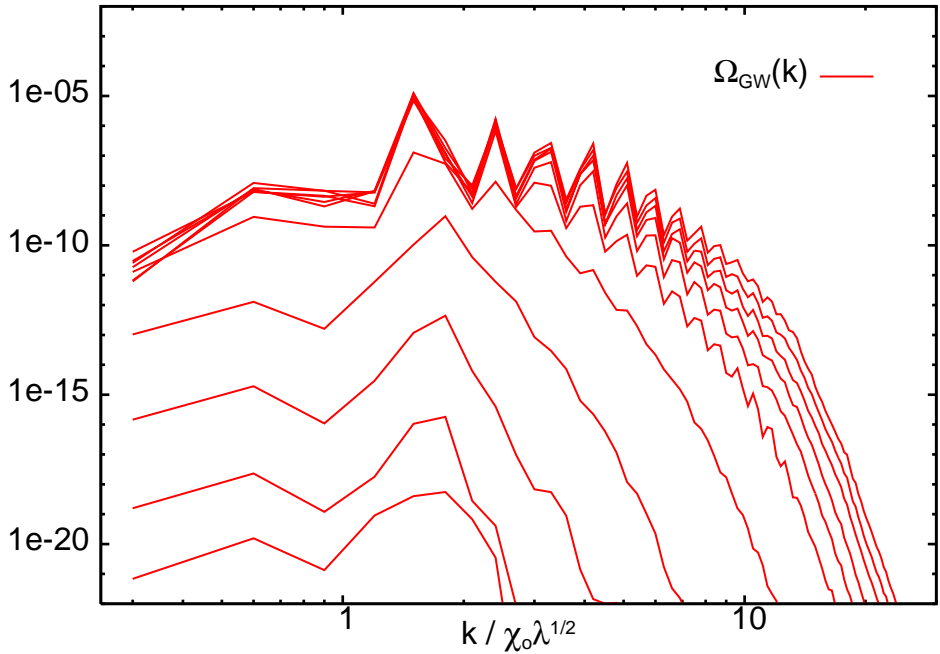}
\end{center}
\vspace*{-5mm}
\caption{The spectrum of the gravitational waves' energy density, 
for the pure case, with $\lambda = 10^{-14}$. Again, we show 
the spectrum accumulated up to different times during GW 
production, normalized to the total instant density at each time.
The plot corresponds to a N = 128 lattice simulation, 
from $\tau = 0$ to $\tau = 2000$.}
\label{fig7}
\vspace*{-3mm}
\end{figure}

We begin our simulations at the
end of inflation, when the homogeneous inflaton verifies $\chi_0
\approx 0.342 M_p$ and $\dot\chi_0 \approx 0$. We took initial quantum
(conformal) fluctuations $1/\sqrt{2k}$ for all the modes up to a
certain cut-off, and only added an initial zero-mode for the inflaton,
$\chi_c(0) = 1$, $\chi_c(0)' = 0$. In Figs.~\ref{fig6} and~\ref{fig7}, 
we show the 
evolution of $\Omega_{_{\rm GW}}$ during reheating, normalized to the 
instant density at each time step, for the coupled and the pure 
case, respectively. In the case with an extra scalar field, 
the amplitude of the GWB saturates at the end of parametric 
resonance, when the fields 
variances have been stabilized. This is the beginnig of the turbulent 
stage in the scalar fields, which seems not to source anymore the 
production of GWs, as already stated in Refs.\cite{EastherLim,DufauxGW}.
For the pure case, we also see the saturation of the amplitude of the 
spectra, see Fig.~\ref{fig7}, although the high momentum tail seems to 
slightly move toward higher values.

%%%%%%%%%%%%%%%%%%%%%%%%%
%%%%%%%%%%%%%%%%%%%%%%%%%

Of course, in either case, with and without
an extra field $\phi$, in order to predict today's spectral window of
the GW spectrum, we have first to normalize their energy density
at the end of GW production to the total
energy density at that moment; then to redshift the GW
spectra from that moment of reheating, taking into account that the
rate of expansion have changed significantly since the end of
inflation, see Eq.(\ref{redshift}). In particular, the shape and 
amplitude of GW spectra for the case with the extra scalar field
coupled to the inflaton with $q = 120$, seems to coincide 
with the espectra shown in Ref.~\cite{DufauxGW}. 
On the other hand, we also reproduce %in Fig.~\ref{fig5} 
a similar spectra to the one shown in~\cite{TkachevGW}, 
for the case of the pure quartic model. % with $\lambda = 10^{-14}$.  
Thanks to the tremendous gain in
computer power, we were able to resolve the 'spiky' pattern
of that spectrum with great resolution. For the first time,
it is clearly observed the exponential tail for large frequencies, 
%see Fig.~\ref{fig7}, 
not shown in Ref.~\cite{TkachevGW}. The most
remarkable fact, is that we also confirm that the peak structure in
the GW power spectrum, see Fig.~\ref{fig7}, remains clearly visible at
times much later than the one at which those peaks have dissapeared in
the scalar fields' power spectrum. So, as pointed out in
Ref.~\cite{TkachevGW}, this characteristic feature will allow us to
distinguish this particular model from any other.

Let us emphasize that we have run the simulations till times 
much greater than that of the end of the resonance stage, both for 
the pure and the coupled case. The role of the
turbulence period after preheating seems, therefore, not to be very 
important, despite its long duration. Apparently, the \textit{no-go} 
theorem about the suppresion of GW at turbulence, discussed in~\cite{DufauxGW}, 
is fulfilled. In Refs.~\cite{CEWB,Dufaux} it was 
pointed out that gauge couplings or trilinear interactions could be 
responsible for a fast thermalization of the universe after inflation 
(see also Ref.~\cite{FK}), but as long as this takes place after the end 
of the resonace stage, in principle this should not affect the results 
shown above.

%\begin{widetext}

%%%%%%%%%%%%%%%%%%%%%%%%%
\begin{figure}[h]
\vspace{-8mm}
\begin{center}
\includegraphics[width=12cm,angle=-90]{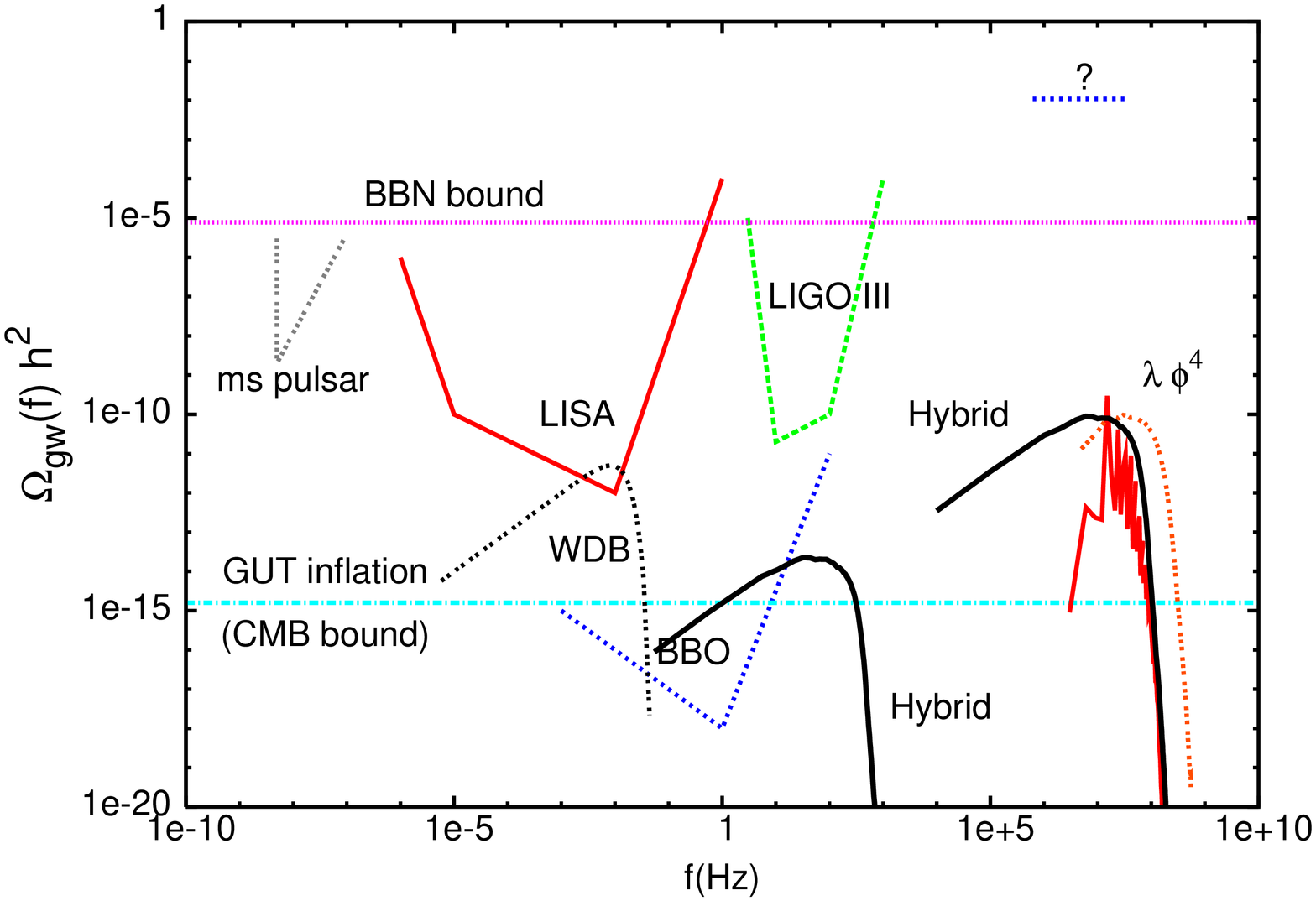}
\end{center}
\vspace*{-2cm}
\caption{The sensitivity of the different gravitational wave
experiments, present and future, compared with the possible stochastic
backgrounds; we include the White Dwarf Binaries (WDB)~\cite{WDB} and
chaotic preheating ($\lambda\phi^4$, coupled and pure) for comparison.
Note the two well differentiated backgrounds from high-scale and
low-scale hybrid inflation. The bound marked (?) is estimated from 
ultra high frequency laser interferometers' 
expectations~\cite{Nishizawa2007}.}
\label{fig5}
\vspace*{-3mm}
\end{figure}
%%%%%%%%%%%%%%%%%%%%%%%%%

%\end{widetext}

\section{Conclusions}

To summarize, we have shown that hybrid models are very efficient
generators of gravitational waves at preheating, in three well defined
stages, first via the tachyonic growth of Higgs modes, whose gradients
act as sources of gravity waves; then via the collisions of highly
relativistic bubble-like structures with large amounts of energy
density, and finally via the turbulent regime (although this effect
does not seem to be very significant in the presence of scalar
sources), which drives the system towards thermalization. These
waves remain decoupled since the moment of their production, and thus
the predicted amplitude and shape of the gravitational wave spectrum
today can be used as a probe of the reheating period in the very early
universe. The characteristic spectrum can be used to distinguish
between this stochastic background and others, like those arising from
NS-NS and BH-BH coalescence, which are decreasing with frequency, or
those arising from inflation, that are flat~\cite{SKC}.
\\

We have plotted in Fig.~\ref{fig5} the sensitivity of planned GW
interferometers like LIGO, LISA and BBO, together with the present
bounds from CMB anisotropies (GUT inflation), from Big Bang
Nucleosynthesis (BBN) and from milisecond pulsars (ms pulsar).  Also
shown are the expected stochastic backgrounds of chaotic inflation
models like $\lambda\phi^4$, both coupled and pure, as well as
the predicted background from two different hybrid inflation models, a
high-scale model, with $v=10^{-3}M_P$ and $\lambda\sim g^2\sim0.1$,
and a low-scale model, with $v=10^{-5}M_P$ and $\lambda\sim g^2\sim
10^{-14}$, corresponding to a rate of expansion $H\sim 100$ GeV. The
high-scale hybrid model produces typically as much gravitational waves
from preheating as the chaotic inflation models. The advantage of
low-scale hybrid models of inflation is that the background produced
is within reach of future GW detectors like BBO~\cite{BBO}. It is
speculated that future high frequency laser interferometers could
be sensitive to a GWB in the MHz region~\cite{Nishizawa2007}, although
they are still far from the bound marked with an interrogation sign.

For a high-scale model of inflation, we may never see the predicted GW
background coming from preheating, in spite of its large amplitude,
because it appears at very high frequencies, where no detector has yet
shown to be sufficiently sensitive, unless the spectrum can be 
extrapolated to lower frequencies, where there are interferometric 
detectors like BBO which could see a signal. On the other hand, if inflation
occured at low scales, even though we will never have a chance to
detect the GW produced during inflation in the polarization
anisotropies of the CMB, we do expect gravitational waves from
preheating to contribute with an important background in sensitive
detectors like BBO. The detection and characterization of such a GW
background, coming from the complicated and mostly unknown epoch of
rehating of the universe, may open a new window into the very early
universe, while providing a new test on inflationary cosmology.
\\

\section*{Acknowledgments} 
We wish to thank Alfonso Sastre for a very fruitful collaboration.
JGB thanks the organisers of the Japan GRG 17th Meeting in Nagoya
for a very enjoyable conference and legendary hospitality. This work is
supported in part by CICYT projects FPA2003-03801 and FPA2006-05807,
by EU network "UniverseNet" MRTN-CT-2006-035863 and by CAM project
HEPHACOS S-0505/ESP-0346. D.G.F. and A.S. acknowledges support from a
FPU-Fellowship from the Spanish M.E.C.  We also acknowledge use of the
MareNostrum Supercomputer under project AECT-2007-1-0005.

\end{document}